\documentclass[3p,authoryear]{elsarticle}

\newcommand{\edit}[2]{\sout{#1}\textcolor{blue}{#2}}
\newcommand{\edittwo}[2]{\textcolor{violet}{\sout{#1}#2}}
\newcommand{\editthree}[2]{\textcolor{blue}{\sout{#1}#2}}
\newcommand{\turnoffedits}{\renewcommand{\edit}[2]{##2}\renewcommand{\edittwo}[2]{##2}\renewcommand{\editthree}[2]{##2}}

\usepackage{graphicx}
\usepackage{amssymb}
\usepackage{gensymb}
\usepackage{lineno}
\usepackage{chemformula}
\usepackage{booktabs}
\usepackage{subfig}
\usepackage[normalem]{ulem}
\usepackage{pdflscape}
\usepackage{afterpage}
\usepackage{capt-of}
\usepackage{multirow}
\usepackage{tablefootnote}
\usepackage{array}
\usepackage{adjustbox}
\usepackage{color}

\graphicspath{{figures/}}
\newcommand{\Ncases}{762}
\journal{Physics of the Earth and Planetary Interiors}
\begin{document}
\turnoffedits 

\begin{frontmatter}
\title{Low volcanic outgassing rates for a stagnant lid Archean Earth with graphite-saturated magmas}
\author[fub,cam]{Claire Marie Guimond\corref{cor}}\ead{cmg76@cam.ac.uk}
\author[fub]{Lena Noack}
\author[fub,dlr]{Gianluigi Ortenzi}
\author[dlr]{Frank Sohl}
\address[fub]{Freie Universit\"at Berlin, Institute of Geological Sciences, Malteserstrasse 74-100, 12249 Berlin, Germany}
\address[cam]{Bullard Laboratories, Department of Earth Sciences, University of Cambridge, Madingley Rise, Cambridge CB3 0EZ, UK}
\address[dlr]{German Aerospace Centre, Institute of Planetary Research, Rutherfordstrasse 2, 12489 Berlin, Germany}
\cortext[cor]{Corresponding author}

\begin{abstract}
Volcanic gases supplied a large part of Earth's early atmosphere, but constraints on their flux are scarce. Here we model how C-O-H outgassing could have evolved through the late Hadean and early Archean, under the conditions that global plate tectonics had not yet initiated, all outgassing was subaerial, and graphite was the stable carbon phase in the melt source regions. The model fully couples numerical mantle convection, partitioning of volatiles into the melt, and chemical speciation in the gas phase. The mantle oxidation state (which may not have reached late Archean values in the Hadean) is the dominant control on individual species' outgassing rates because it affects both the carbon content of basaltic magmas and the speciation of degassed volatiles. Volcanic gas from mantles more reduced than the iron-w\"ustite mineral redox buffer would contain virtually no CO$_2$ because (i) carbonate ions dissolve in magmas only in very limited amounts, and (ii) almost all degassed carbon takes the form of CO instead of CO$_2$. For oxidised mantles near the quartz-fayalite-magnetite buffer, we predict median CO$_2$ outgassing rates of less than approximately 5 Tmol yr$^{-1}$, still lower than the outgassing rates used in many Archean climate studies. Relatively weak outgassing is due in part to the redox-limited CO$_2$ contents of graphite-saturated melts, and also to a stagnant lid regime's inefficient replenishment of upper mantle volatiles. Our results point to certain chemical and geodynamic prerequisites for sustaining a clement climate with a volcanic greenhouse under the Faint Young Sun.

\end{abstract}


\end{frontmatter}

\section{Introduction}

The Hadean-Archean atmosphere cloaked the only environment in the universe known to support the origin of life. This atmosphere would have been fed by volatile chemicals outgassed from the interior \citep{Holland1984, Kasting1993, Gaillard2014}, as well as extraterrestrial impacts \citep{Zahnle2020} and perhaps a lingering, degassed magma ocean atmosphere \citep{Katyal2020}. Yet hardly any observable record of the Archean atmosphere remains. Our knowledge surrounding its makeup rests on modelling. Mantle outgassing models lend boundary conditions to the processes controlling the atmospheric composition and planetary climate. However, predictions from these models are very much entwined with certain assumptions---in particular, the tectonic regime and oxidation state of the mantle. Archean outgassing scenarios hinging on mobile plates and an oxidised mantle form the basis of contemporaneous climate studies \citep[e.g.,][]{Sleep2001}. This work offers an alternative to these recurring premises: a fully-coupled numerical estimation of the volcanic outgassing rates of \ch{CO2}, \ch{CO}, \ch{H2O}, and \ch{H2}, for a stagnant lid Earth under a range of mantle oxidation states. 

The oxidation state of the upper mantle controls the chemical speciation of volcanic gas. Increasing the mantle's oxygen fugacity, $f_{\rm O_2}$, raises the ratio of oxidised to reduced volatile species in the melt phase: [\ch{CO2}]/[\ch{CO}] and [\ch{H2O}]/[\ch{H2}] in the C-O-H system \citep{Holland1984}. As the melt rises adiabatically through conduits, its $f_{\rm O_2}$ re-equilibriates with the decreasing pressure; gas abundances at the surface are also linked to the mantle source region \citep{Kasting1993}. On modern Earth, for example, the dominant species in volcanic gas are \ch{H2O} and \ch{CO2}, in line with a mantle source $f_{\rm{O_2}}$ in equilibrium with the quartz-fayalite-magnetite (QFM) mineral buffer \citep{Holland1984}. 

Besides the effect of $f_{\rm{O_2}}$ on volatile speciation, there can be a compounding redox control on the partitioning of carbon into the melt. This effect becomes relevant when graphite is the stable carbon phase in the upper mantle. During melting, carbonate ions are produced from a reaction of graphite with oxygen via \ch{CO2} \citep{Holloway1992}. The total carbon outgassing flux depends on how much carbon (as \ch{CO3^{2-}}) can dissolve into the melt. 

Although these two redox effects are well-known, the value itself of the Archean mantle $f_{\rm O_2}$ remains weakly constrained at best. Immediately after core formation, the magma ocean must have been reduced \citep{Wood2006}. Meanwhile, scant geological evidence suggests that by $\sim$3.8~Ga at the latest, the uppermost mantle was already about as oxidised as today \citep{Canil1997, Delano2001, Trail2011, NICKLAS2019, Armstrong2019}. Yet uncertainty persists on the timing of this oxidation (see section \ref{sec:redox}). Therefore, to cover the possible spread of redox states, we investigate outgassing fluxes under a wide range of mantle $f_{\rm O_2}$, especially in proportion to other unknown parameters such as the mantle volatile content and thermal state.

\editthree{Our study considers}{Whilst this study is set up to explore many redox scenarios, it will be limited to one tectonic regime. For this we consider} the end-member case of stagnant lid convection \citep{Debaille2013}. The possibility of a stagnant lid regime appears underexplored in early Earth climate studies, despite the ambiguity around plate tectonics' operation through the Archean \citep[e.g.,][]{Brown2020}\editthree{}{; as such we hope to complement other work assuming a mobile lid}. This modelling decision is important because mobile lids may be associated with much higher outgassing rates than stagnant lids \citep{noack2014can, Noack2017}. The mechanistic explanations are not straightforward---being a combination of competing effects discussed later in section \ref{sec:stagnant-lid}---and would benefit from dedicated work in the future.

By numerically coupling mantle convection, redox-dependent melt partitioning of volatiles, and redox-dependent speciation, and by investigating a broader range of mantle oxidation states, we can advance the current literature dedicated to C-O-H outgassing from solid-state mantles of rocky planets in the solar system and elsewhere \citep{ONeill2007,Grott2011,noack2014can,Noack2017,tosi2017habitability,dorn2018outgassing, Ortenzi2020}. Previous studies have tended to assume an oxidising mantle, akin to that of present-day Earth---exceptions include \citet{Grott2011} for the reduced early Mars, and \citet{tosi2017habitability}, \citet{Ortenzi2020}, and \citet{Liggins2020} for fiducial rocky exoplanets. 
Further, numerical convection carries the advantage of resolving the local temperatures and pressures of melt parcels, to which $f_{\rm O_2}$ is very sensitive. 

An approach parallel to the one taken here is to parameterise outgassing fluxes in terms of melt production \citep[e.g.,][]{Sleep2001, Papuc2008, kite2009geodynamics, Kadoya2015, Foley2016, CHARNAY2017, KT2018, Foley2018, Krissansen-Totton2020}. In parameterised models, mantle melting rates are often scaled to an estimate of the interior heat flux. Past outgassing rates can then be calibrated to present-day estimates. \editthree{Because}{These types of outgassing parameterisations may imply higher outgassing fluxes than do numerical convection models for two reasons. Firstly, heat flow scalings \editthree{}{extrapolated backwards in time }will necessarily lead to higher-than-modern outgassing rates because} parameterised thermal history models predict high heat flow early in a planet's evolution \citep[e.g.,][]{TURCOTTE1980}. \editthree{This can be misleading for two reasons. First}{Secondly}, 1D models tend to overestimate melt fractions early in a planet's thermal evolution, since they cannot spatially resolve depletion upon partial melting---local depletion would inhibit further local melting for the same temperature. \editthree{Secondly, heat flow considerations alone may underestimate the importance of other factors which control the melt concentrations and speciation of volatiles. Thirdly, a given scaling extrapolated backwards in time would assume no change in tectonic regime.}{In addition to these systematic effects, applying a single scaling relationship assumes no change in mantle oxidation state or tectonic regime.}

Planetary outgassing rates require attention because they lie at the core of at least two unsolved Archean mysteries. The first is that reduced gases such as \ch{H2} and \ch{CO} may be necessary for prebiotic chemistry. The production of HCN requires a reducing atmosphere with C/O $\geq$ 1, for example \citep{Rimmer2019}. However, whether \ch{H2} and \ch{CO} could have comprised a significant part of volcanic gas at the time remains debated \citep[e.g.,][]{Zahnle2020}. The second mystery is that sufficiently large partial pressures of greenhouse gases seem required to incubate early Earth under lower solar luminosity \citep{charnay2020} and make sense of liquid surface water \citep{Sagan1972}. Although no solution to this paradox has emerged as the likely favourite, most atmospheric evolution scenarios proposed so far rely on \ch{CO2} outgassing fluxes at least as high as the present day \citep[e.g.,][]{Sleep2001, Wordsworth2013, KANZAKI2018}. Neither of these two driving questions may ever be resolvable without definite constraints on the timing of mantle oxidation, or plate tectonics' initiation. Nevertheless, this study, by more rigorously linking stagnant lid outgassing fluxes to the unknown oxidation state, might help contextualise these problems within Earth system evolution.

\section{Methods}

We trace the movement of volatiles through the mantle to the atmosphere for a stagnant-lid Earth, from just after the moon-forming impact to around the first rock record. Our model is similar to the one described in \citet{Ortenzi2020}, who additionally calculated gas solubility in the melt (relevant for the higher outgassing pressures expected on exoplanets greater than an Earth mass), but coupled partitioning and volatile speciation \textit{a posteriori} within the convection simulations, rather than directly.

To simulate mantle dynamics, we use a 2D convection model, which is detailed in \citet{noack2016modeling}. The heavily-benchmarked model couples solid-state convection in the mantle with spatially-resolved melt fractions and volatile partitioning in the melt \citep{Holloway1992, Katz2003, Grott2011}. This is further coupled with a chemical speciation model \citep{French1966, Holloway1981, Fegley2013, gaillard2014theoretical, schaefer2017redox}. At each time step the model produces outgassed masses of \ch{H2O}, \ch{H2}, \ch{CO2}, and \ch{CO}. We simulate 700~Myr of convection, with time-zero nominally corresponding to 4.5~Ga.

\subsection{Numerical convection model}

The convection code solves the conservation equations of mass, momentum and energy in the rocky mantle by using a particle-in-cell method. We use the 2D spherical annulus method in \citet{HERNLUND2008} to divide the mantle into a mesh of cells with \editthree{uniform height of 25 km}{an average lateral and radial resolution of 50 km between grid points. The radial resolution coarsens linearly with depth, varying from 10 km at the surface to 90 km at the core-mantle boundary}. To reduce computational costs, we model one quarter of the 2D annulus, corresponding to 307 cells in the lateral direction. \editthree{}{We found that minor changes in resolution did not affect degassing.}

We model compressible convection based on the truncated anelastic liquid approximation (TALA), using pre-calculated depth-dependent profiles for thermodynamic parameters such as the mineral-dependent density, the thermal expansion coefficient and the heat capacity for an adiabatic temperature profile. We neglect any possible initiation of plate tectonics or other surface recycling mechanisms and instead allow for a stagnant lid to form. 

We take into account time-dependent radiogenic heating and heat loss from the core. The radiogenic heat sources decline over time and would reach present-day Earth values of 79.5 ppb Th, 240 ppm K and 20.3 ppb U after 4.5 Gyr of thermal evolution, following \citet{schubert_turcotte_olson_2001}. We did not include the partitioning of heat sources in the crust, and instead applied homogeneous radiogenic heat sources over the entire domain.

We assume a hydrated pyrolitic mantle with an Arrhenius viscosity, $\eta$, that depends on the local temperature $T$ and pressure $p$, following \citet{Karato1993}:
\begin{equation}
    \eta(T,p) = \frac{1}{2 A} d^r \exp\left(\frac{E+pV}{RT}\right),
\end{equation}
with, for the upper mantle, a prefactor $A=3.7 \times 10^{-19}$ m$^{-r}$ Pa$^{-1}$ s$^{-1}$, grain size $d=1$~mm, grain size exponent $r=2.5$, activation energy $E=240$ kJ mol$^{-1}$, and activation volume $V=5.0$ cm$^{3}$ mol$^{-1}$, where $R$ is the gas constant. The lower mantle rheology behaves as in \citet{Noack2017}, with a viscosity prefactor and pressure-dependent activation enthalpy following \citet{Tackley2013}\editthree{}{; that is, $V$ decreases with pressure in the lower mantle}. \editthree{In the \citeauthor{Tackley2013} study, however, the viscosity was increased by a factor of 100 to allow for numerically-feasible simulations. Here, we do not artificially scale $\eta$, but instead limit it to a minimum of $10^{21}$~Pa~s to ensure numerical stability in the mantle for all investigated parameter combinations.}{Whereas \citet{Tackley2013} increased their viscosities by a factor of 100 to allow for numerically-feasible simulations, we do not also adopt this scale factor, so the lower mantle is less viscous compared to their study.}

\editthree{}{Whilst we model a wet mantle, we do not consider local viscosity variations due to heterogeneous water contents. }We do not implement an aesthenosphere, and may underestimate the convective strength locally in the upper mantle. \editthree{Our initial Rayleigh number is therefore around $\sim$5 $\times 10^7$, depending on the initial temperature profile.}{Our initial Rayleigh number, estimated for a global average viscosity of about $10^{21}$ Pa~s, is around $5\times10^7$. However, depending on the initial temperature profile and its local variations, the viscosity can be much lower than this---especially in the upper mantle---permitting more vigorous convection in some regions.}

\subsection{Melting model}
The petrological model simulates partial melting of the silicate mantle. Melt forms if the local temperature is greater than the mantle solidus temperature, $T_{\rm solidus}$, at that temperature and pressure. The local melt fraction, $F$, is
\begin{equation}\label{eq:melt-fraction}
    F = \frac{T - T_{\rm solidus}}{T_{\rm liquidus} - T_{\rm solidus}}.
\end{equation}
The solidus and liquidus parameterisations, given in \citet{noack2014can}, are fits to experimental data for peridotite melting valid up to 15~GPa \citep{smet1999evolution}. The liquidus temperature, $T_{\rm liquidus}$, is constant for a given pressure. The solidus temperature decreases with water content in the melt \citep{Katz2003}, and increases upon depletion due to prior melting. \editthree{}{Melting also consumes latent heat, feeding back into the local mantle temperatures. }We assume that the upper mantle melt is buoyant only at pressures of less than 12~GPa \citep{Ohtani1995}, hence we neglect any contribution from melting at greater depths. 

The local melt fractions add up to give the total chemical depletion of the mantle. Here, depletion is adapted to the fact that dehydration increases the solidus temperature. We recalculate the depletion at each time step using the updated solidus temperature. Upon 30\% depletion of the primordial peridotite mantle, the residual mantle therefore always resembles the harzburgite composition, independently of if it having started wet or dry. Depletion is limited to 30\%.\editthree{}{ We do not consider density variations in the mantle residue upon melting; that is, there is no buoyancy effect due to depletion.}

\subsection{Volatile partitioning model}\label{sec:partitioning-methods}

At each time step, if the local melt fraction is greater than zero, we calculate the amounts of C and \ch{H2O} that enter the melt phase, and subtract them from the local chemical inventories. These inventories (as well as other local information such as the depletion) are traced via mass-independent particles that flow along convective stream lines. We ensure that each cell, which is fixed in space, has between 3 and 8 particles. No local partitioning occurs once a particle's volatile inventory is emptied; if a particle is dehydrated, it remains so. As the mantle convects, however, cells can be replenished by fresh particles rising up from the lower mantle, which replace the dehydrated particles. 

Carbon dissolves into the melt as \ch{CO3^2-}; its mole fraction, $X_{\ch{CO3^2-}}^{\rm melt}$, depends strongly on $f_{\rm O_2}$:
\begin{equation}\label{eq:c_partition}
    X_{\ch{CO3^2-}}^{\rm melt} = \frac{K_{\rm I} \, K_{\rm II} \, f_{\ch{O2}}}{1+K_{\rm I} \, K_{\rm II} \, f_{\ch{O2}}},
\end{equation}
where $K_{\rm I}$ and $K_{\rm II}$ are the equilibrium constants governing, respectively, the formation of \ch{CO2} from graphite and \ch{CO3^{2-}} from \ch{CO2},
\begin{align}
    \log_{10} K_{\rm I} &= a - b \, T + c \, T^2 + d \, \frac{p-1}{T}\\ 
    \log_{10} K_{\rm II} &= e - \frac{f}{T} - g \frac{p-1000}{T},
\end{align}
where $a = 40.07639$, $b = 2.53932\times 10^{-2}$, $c = 5.27096\times 10^{-6}$, $d=0.0267$, $e = -6.24763$, $f=282.56$, $g=0.119242$, $T$ is in K, and $p$ is in bar \citep{Holloway1992, Grott2011, Hirschmann2008}. 

The amount of \ch{CO2} in weight percent, here denoted $\chi$ to distinguish from mole percent, is calculated from $X_{\ch{CO3^2-}}^{\rm melt}$, assuming thermodynamic equilibrium:
\begin{equation}\label{eq:c_partition2}
    \chi_{\ch{CO2}}^{\rm melt} = \left. \left[\frac{M_{\ch{CO2}}}{\rm FWM} \, X_{\ch{CO3^2-}}^{\rm melt} \right] \; \middle/ \; \left[ 1 - \left(1-\frac{M_{\ch{CO2}}}{\rm FWM} \right) X_{\ch{CO3^2-}}^{\rm melt}\right] \right. ,
\end{equation}
where $M_{\ch{CO2}}$ is the molar mass of carbon dioxide in g~mol$^{-1}$ and FWM is the formula weight of the melt normalised to one oxygen, equal to 36.594~g~mol$^{-1}$ for an example tholeiitic basaltic \citep{Holloway1992, Grott2011}. The formula weight typically varies only by few percent for different basalt compositions. \editthree{}{Note that although (\ref{eq:c_partition}--\ref{eq:c_partition2}) do not directly involve the mantle source concentration, $\chi_{\ch{CO2}}^{\rm melt}$ is capped at the source concentration locally.}

We assume that all solid carbon exists as graphite, since we only consider shallow melting below 12~GPa. This assumption may not be valid for the most oxidised cases (IW~+~4), as carbonate may become the stable carbon species for these pressures and $f_{\rm O_2}$ \citep{Stagno2019}. However, in such oxidised conditions we already know that \ch{CO2} is outgassed, and neglect any carbonate stability.

For water partitioning, we use the batch melting formula:
\begin{equation}\label{eq:h_partition}
    \chi_{\ch{H2O}}^{\rm melt} = \frac{\chi_{\ch{H2O}}^{\rm rock}}{D_{\ch{H2O}}+F(1-D_{\ch{H2O}})},
\end{equation}
where $\chi_{\ch{H2O}}^{\rm rock}$ is the local weight fraction of water in the solid mantle, $F$ is the local average melt fraction, and $D_{\ch{H2O}}$ is a partition coefficient, which we set at 0.01 based on the Ce partitioning coefficient \citep{MICHAEL1995, Katz2003}.

Whilst the melt rises adiabatically towards the surface, we only allow a fixed fraction of the melt, $\chi_{\rm extr}$, to reach the surface as extrusive volcanism. Only this extrusive melt contributes to outgassing into the atmosphere. The mantle volatile concentrations are small enough to not saturate \citep{Katz2003}. We do not consider submarine degassing, so virtually all gas molecules enter the atmosphere at these lower surface pressures. Our model therefore assumes complete outgassing of the melt that reaches the surface, with no residual volatiles remaining in the melt. Finally, we do not consider any possible re-equilibration or contamination as material rises through the lithosphere, and neglect whether the melt interacts with the surrounding rock or with fluids at higher pressures.

\subsection{Volatile speciation model}\label{sec:speciation-methods}

\begin{table}
\caption{Thermodynamic parameters for basaltic melt.
\label{table:thermo_param}}  
\centering   
\begin{tabular}{l p{5cm} r l l}
\hline\hline     
\noalign{\vskip 1mm}   
 \textsc{Symbol} & \textsc{Description} & \textsc{Value} & \textsc{Units} & \textsc{Reference} \\    
\hline     
\noalign{\vskip 1mm}   
   $\alpha_{\rm melt}$ & Thermal expansion coefficient & $3\times10^{-5}$ & K$^{-1}$ & \citet{Afonso2005} \\     
   $\rho_{\rm melt}$ & Density &  3000 & kg m$^{-3}$ & \citet{Lesher2015}\\
   $C_{p,{\rm melt}}$ & Heat capacity &  1793 & J kg$^{-1}$ K$^{-1}$ & \citet{Lesher2015} \\
\hline                                   
\end{tabular}
\end{table}

We retrieve the masses of outgassed volatiles using the ``Equilibrium constants and mass balance method" first appearing in \citet{French1966}, and used widely in other studies \citep{Holland1984, Gaillard2014, Fegley2013, schaefer2017redox, Ortenzi2020}. The [H$_2$]/[H$_2$O] and [CO]/[CO$_2$] molar ratios are governed by the chemical equilibria,
\begin{equation}\label{equilibrium_H2O}
    \ch{2 H2 +  O2 <=> 2 H2O},
\end{equation}
\begin{equation}\label{equilibrium_CO2}
    \ch{CO + (1/2) O2 <=> CO2},
    \end{equation}
\edit{i.e., proportional to $f_{\rm O_2}^{1/2}$, }{}assuming that the volcanic gases are in equilibrium with their magmas. These chemical reactions have respective equilibrium constants $K_{\rm III}$ and $K_{\rm IV}$, which are calculated as follows:
\begin{equation}\label{eq:equilibrium_constant_hydrogen}
   {K_{\rm III}}  = \exp\left(\frac{-\Delta_rG^0_{(\ref{equilibrium_H2O})} }{RT}\right) = \frac{\left(X_{\ch{H2O}}\right)^2}{\left(X_{\ch{H2}}\right)^2f_{\ch{O2}}} , 
\end{equation}
\begin{equation}\label{eq:equilibrium_constant_carbon}
    {K_{\rm IV}}  = \exp\left(\frac{-\Delta_rG^0_{(\ref{equilibrium_CO2})}}{RT}\right) = \frac{X_{\ch{CO2}}}{X_{\ch{CO}}}\frac{1}{\left(f_{\ch{O2}}\right)^{1/2}},
\end{equation}
where $R$ is the universal gas constant (8.314 J K$^{-1}$ mol$^{-1}$), $X_{i}$ is the mole fraction of species $i$, and $\Delta_rG^0_{(\ref{equilibrium_H2O})}$ and $\Delta_rG^0_{(\ref{equilibrium_CO2})}$ are the Gibbs free energies of reaction in J mol$^{-1}$ for (\ref{equilibrium_H2O}) and (\ref{equilibrium_CO2}) respectively---functions of temperature given in \citet{Ortenzi2020}, based on \citet{Fegley2013}. Equations (\ref{eq:equilibrium_constant_hydrogen}) and (\ref{eq:equilibrium_constant_carbon}) assume surface pressure $p_{\rm surf} = 1$~bar, whilst $T$ is taken to be the temperature the melt would have once it has risen adiabatically to the surface:
\begin{equation}
    T (p = p_{\rm surf}) = T_{\rm melt}(p_{\rm melt}) \cdot \exp \left[ -\frac{\alpha_{\rm melt}}{\rho_{\rm melt} \; C_{p,{\rm melt}}} (p_{\rm melt} - p) \right],
\end{equation}
where $C_{p,{\rm melt}}$ is the heat capacity of the melt in J~kg$^{-1}$~K$^{-1}$, $\alpha_{\rm melt}$ is the thermal expansion coefficient in K$^{-1}$, $\rho_{\rm melt}$ is the density in kg~m$^{-3}$, and $T_{\rm melt}$ and $p_{\rm melt}$ are the local temperature in K and pressure in Pa of the melt source. Values for the thermodynamic parameters are listed in Table \ref{table:thermo_param}.

From the model outlined above, we can calculate the outgassed mass $M_i$ in kg of each volatile species $i$ at a given time, accumulated over all previous time steps and over all cells in the domain:
{
    \begin{equation}\begin{split}
    M_{\ch{H2O}}^{\rm atm} = &\chi_{\rm extr} \sum\limits_{j=1}^{\rm times} \sum\limits_{k=1}^{\rm cells} \chi_{\ch{H2O}, j, k}^{\rm melt} F_{j,k} V_{k} \frac{X_{\ch{H2O}}}{X_{\ch{H2}}+X_{\ch{H2O}}} \rho_{\rm mtl}
    \\
    M_{\ch{H2}}^{\rm atm} = &\chi_{\rm extr} \sum\limits_{j=1}^{\rm times} \sum\limits_{k=1}^{\rm cells} \chi_{\ch{H2O}, j, k}^{\rm melt} F_{j,k} V_{k} \frac{X_{\ch{H2}}}{X_{\ch{H2}}+X_{\ch{H2O}}} \frac{m_{\ch{H2}}}{m_{\ch{H2O}}} \rho_{\rm mtl}
    \\
    M_{\ch{CO2}}^{\rm atm} = &\chi_{\rm extr} \sum\limits_{j=1}^{\rm times} \sum\limits_{k=1}^{\rm cells} \chi_{\ch{CO2}, j, k}^{\rm melt} F_{j,k} V_{k} \frac{X_{\ch{CO2}}}{X_{\ch{CO}}+X_{\ch{CO2}}}\rho_{\rm mtl}
    \\
    M_{\ch{CO}}^{\rm atm} = &\chi_{\rm extr} \sum\limits_{j=1}^{\rm times} \sum\limits_{k=1}^{\rm cells} \chi_{\ch{CO2}, j, k}^{\rm melt} F_{j,k} V_{k} \frac{X_{\ch{CO}}}{X_{\ch{CO}}+X_{\ch{CO2}}} \frac{m_{\ch{CO}}}{m_{\ch{CO2}}} \rho_{\rm mtl},
\end{split}\label{eq:m_atm}
\end{equation}}
where $\chi_{\rm extr}$ is the percentage of volcanism that is extrusive, $F_{j,k,}$ is the local melt fraction in a cell, $V_{k}$ is the volume of that cell in m$^{3}$, $m_i$ is molar mass of species $i$ in kg mol$^{-1}$, and $\rho_{\rm mtl}$ is the local density of the solid mantle in kg~m$^{-3}$. Note that $\chi$ has units of weight fraction and $X$ has units of mole fraction. The melt volumes correspond to a 2D annulus, later scaled up to 3D by a dimensionalisation factor.  

The partial pressure in Pa outgassed to the atmosphere for species $i$ is: 
\begin{equation} \label{eq:partpresh}
p _i= \frac{M_{\rm tot} X_{i,{\rm tot}} g}{4\pi R_E^2},
\end{equation}
where $M_{\rm tot}$ is the sum over species of the outgassed masses from equation (\ref{eq:m_atm}) in kg, $X_{i, {\rm tot}}$ is the species' mole fraction, $g$ is the acceleration due to gravity at the surface (9.8 ms$^{-2}$), and $R_E$ is the radius of Earth (6371 km). We assume that the gas reaches the atmosphere instantaneously.

\subsection{Initial conditions and free parameters} \label{sec:input-range}

\begin{table*}
\caption{Input parameters varied in this study. Values are drawn from random uniform distributions bounded by the values in the Range column, with the exception of $T_{\rm ini}$ (see text). \label{tab:inputs}}
\centering
\begin{tabular}{l l l r}
\hline
\hline
\textsc{Symbol} & \textsc{Description} & \textsc{Range} & \textsc{Units} \\
\hline
\noalign{\vskip 0.5mm} 
$\log({f_{\rm O_2}}) - {\rm IW}$ & Mantle redox shift from IW 
& [-3, 4] & dex  \\
$\chi_{\rm H_2O}^{\mathrm{ini}}$ & Initial mantle \ch{H2O} content & [50, 450] & wt. ppm \\
$\chi_{\rm CO_2}^{\mathrm{ini}}$ & Initial mantle \ch{CO2} content & [22, 180] & wt. ppm  \\
$T_{\rm surf}$ & Temperature at surface & [273, 333] & K \\
$T_{\rm ini}$ & Initial temperature at top of convecting layer & [1750, \editthree{2100}{2000}] & K \\
$D_{\rm lid}$ & Initial lid thickness & [10, 43] & km  \\
$\Delta T_c$ & Initial core temperature jump & [0, 1750] & K  \\
$\chi_{\rm extr}$ & Extrusive volcanism percentage & [10, 40] & wt. \%  \\
\hline
\end{tabular}
\end{table*}

The main study involves \textgreater 700 cases with random values for eight free input parameters (Table \ref{tab:inputs}), identified as the main unknown parameters potentially affecting the early Earth evolution. Input values for each case are drawn from uniform distributions, with one exception. This section outlines the choice of bounds.

\subsubsection{Choice of initial temperatures}

The initial mantle temperature profile is set by four unknown parameters: planet surface temperature, $T_{\rm surf}$, initial stagnant lid thickness, $D_{\rm lid}$, initial temperature at the base of the lithosphere, $T_{\rm ini}$, and core temperature jump, $\Delta T_c$. Temperatures increase linearly from $T_{\rm surf}$ to $T_{\rm ini}$, and then follow an initially adiabatic profile from below the upper thermal boundary layer to the core-mantle boundary. The temperature contrast between the base of the mantle and the core is fixed at $\Delta T_c$. We enforce an initial temperature cut-off at the solidus. 

Although evidence of surface temperatures covering our age of interest is scant, oxygen isotope ratios from the 4.4~Ga Jack Hills zircons imply a surface temperate enough to host liquid oceans \citep{Valley2002}. Thus we allow $T_{\rm surf}$ to vary from 0 to 60$^\circ$C. 
Our constant $T_{\rm surf}$ is an approximation; in reality this value varies spatially and is linked to outgassing via greenhouse warming. 

The temperatures in Earth's mantle after the last magma ocean stage are unknown. We adopt a range of 1750--\editthree{2100}{2000}~K in $T_{\rm ini}$. \editthree{}{This study's effective distribution of $T_{\rm ini}$ is not uniform because higher values of $T_{\rm ini}$ were more likely to result in numerical instabilities. Values of $T_{\rm ini} > 2000$~K always led to numerical errors. However, temperatures do increase beyond this initial value due to radiogenic heating. }Archean komatiite records suggest an upper mantle potential temperature of at most $\sim$1970 K \citep{HERZBERG2010}, corresponding to upper mantle temperatures comparable with those investigated here (our $T_{\rm ini}$ values refer to the actual temperatures below an initial upper thermal boundary layer, hence the potential temperatures would be slightly smaller). 

The initial adiabatic profile implies an already-convecting mantle. However, at the onset of our simulations, the mantle is gravitationally stable. Hence we slightly perturb the temperature field to trigger convection. Initial temperatures are close to the solidus, so melting and outgassing begin quickly.
Regardless, this temperature profile quickly starts evolving, and its initial shape is not significant to the results. The stagnant lid thickness also adjusts according to the temperature profile. Very thick lithospheres ($\sim$100~km) would inhibit melt production to the extent that no outgassing would occur within 700~Myr.

\subsubsection{Choice of initial volatile contents}

Most of the bulk Earth's accreted volatile content may have been expelled during magma ocean degassing. The crystallised mantle's initial \ch{H2O} and \ch{CO2} contents, $\chi_{\ch{H2O}}^{\rm ini}$ and $\chi_{\ch{CO2}}^{\rm ini}$ in weight percent, are the amounts of volatiles remaining after degassing adopted from the magma ocean model of \citet{ELKINSTANTON2008}. The most water-rich scenario would correspond to just below an Earth ocean mass. These estimates are consistent with more recent modelling work by \citet{Barth2021}. \citet{Hiermajunder2017} found comparable but slightly elevated values considering faster freezing of the magma ocean. \citet{ELKINSTANTON2008} considers primordially-accreted volatile contents of 0.05--0.5 weight percent \ch{H2O} and 0.01--0.1 weight percent \ch{CO2}, although the true values are poorly constrained.

\subsubsection{Choice of mantle oxidation states}\label{sec:redox}

Geochemical clues mostly agree that since around 3.8~Ga or earlier, the Archean upper mantle was about as oxidised as present \citep{Canil1997, Delano2001, Trail2011, NICKLAS2018, Armstrong2019}, although other analyses suggest a more gradual oxidation up to QFM \citep{Aulbach2016, NICKLAS2019}. \citet{Trail2011} investigated Hadean zircons with ages between 4.0--4.36 Ga and derived an oxidation state around QFM. If these zircons did indeed come from primitive mantle melts as suggested by \citet{Trail2011}, this would suggest an oxidised mantle almost directly after the Moon-forming impact. 

We allow mantle $f_{\ch{O2}}$ to vary between the minimum Hadean value of three log-units below the equilibrium with the IW buffer \citep{Wood2006}, and a typical modern value of four log-units above the equilibrium. We assume a constant oxidation state, and later test the effect of $\log(f_{\ch{O2}})$ increasing linearly.

\begin{figure}
    \centering
    \includegraphics[width=0.5\linewidth]{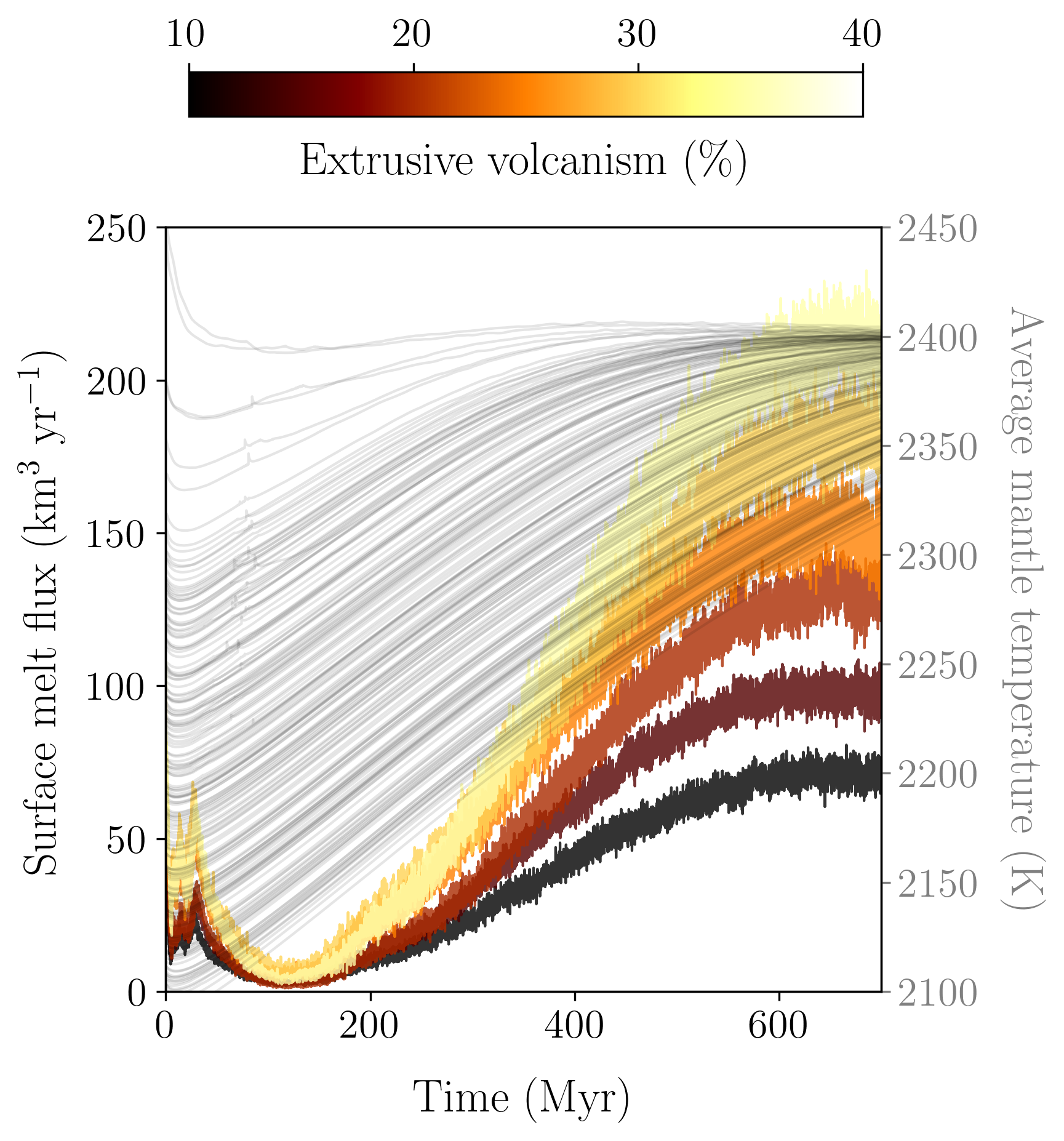}
    \caption{The evolution of the extrusive volumetric melt production rate, assuming a melt density of 3000~kg~m$^{-3}$. Note that these values represent only 10--40\% of the total melt production. Cases are binned to \editthree{10-K increments in initial temperature at the base of the stagnant lid ($T_{\rm{ini}}$)}{5\%-increments in the extrusive volcanism fraction}, and averaged over all other input parameters (Table \ref{tab:inputs}). Lighter colours indicate \editthree{hotter initial conditions}{higher extrusive volcanism fractions}. Overlain in solid grey lines and corresponding to the secondary $y$-axis are the evolutions of the mean temperature over the whole mantle, for each individual model run.}
    \label{fig:melt_vol}
\end{figure}

\subsubsection{Choice of extrusive volcanism percentages}

Only a fraction of magma will rise high enough to extrude and outgas at the surface. The remaining melt in the intrusive component will eventually replenish the mantle. \citet{Crisp1984} estimate an extrusive volcanism percentage of 6--25\% for intracontinental and island arc settings, and 16--33\% at mid-ocean ridges and hotspots. Because this value is unknown for early Earth, we allow it to vary between 10\% and 40\% \citep{Grott2011}. We do not consider any diffuse degassing from intrusive volcanism, although fluid pressure would indeed transport intrusive magmatic gas to the surface through rock fissures.

\section{Results}

Before speciation, the total C (or H) outgassing flux is proportional to the product of the C (or H) melt concentration and the extrusive melt production rate. \editthree{}{In principle, either of these factors has the potential to scale outgassing dramatically, although they do not necessarily show the same variance. We will first present the melt production rates and melt concentrations from the model. Then we will describe the resulting outgassing fluxes, their final speciation modulated by $f_{\ch{O2}}$.}

\subsection{Thermal history and melt production}

The thermal histories underlying the outgassing model are shown in figure \ref{fig:melt_vol}. Mean mantle temperatures gradually increase due to powerful primordial radiogenic heating and relatively sluggish stagnant lid convection. \editthree{Hotter initial temperatures are associated with higher melt production rates.}{The temperature dependence of viscosity acts as a thermostat, such that by 700~Myr, average mantle temperatures converge to $\sim$2320--2400~K.} 

The first $\sim$50~Myr see spikes in the melt flux\editthree{---the magnitude of which depends on $T_{\rm ini}$---followed by a decrease and subsequent settling to a steady value}. Directly after magma ocean solidification, the mantle temperature is still very close to the solidus. In these conditions, slight mass movement from convection can trigger immediate re-melting. This depletes the residual mantle and raises the solidus temperature. Melting is revived once either radiogenic heating warms the upper mantle, or convection upwells undepleted material from the lower mantle to pressures of $\le$12~GPa. 

\editthree{Near steady state, melt is produced at rates of up to 35~km$^3$~yr$^{-1}$. Despite the hotter stagnant-lid mantle with respect to modern Earth, total melt production is similar to \citeauthor{Crisp1984}'s \citeyear{Crisp1984} canonical estimate of 26--34~km$^3$~yr$^{-1}$ for combined magma emplacement and volcanic output (see section \ref{sec:stagnant-lid}).}{}

Note that higher initial temperatures than considered here would lead to unphysical conditions, with most of the mantle above the solidus temperature. Whilst a much wetter mantle might aid melting, we do not find a dramatic effect of $\chi_{\ch{H2O}}^{\rm ini}$ on the overall melt production. \editthree{}{We do not include dehydration's stiffening effect on viscosity.}

\subsection{Melt contents of \ch{H2O} and \ch{CO2}}

\begin{figure}%
    \centering
    \subfloat{{\includegraphics[width=0.45\linewidth]{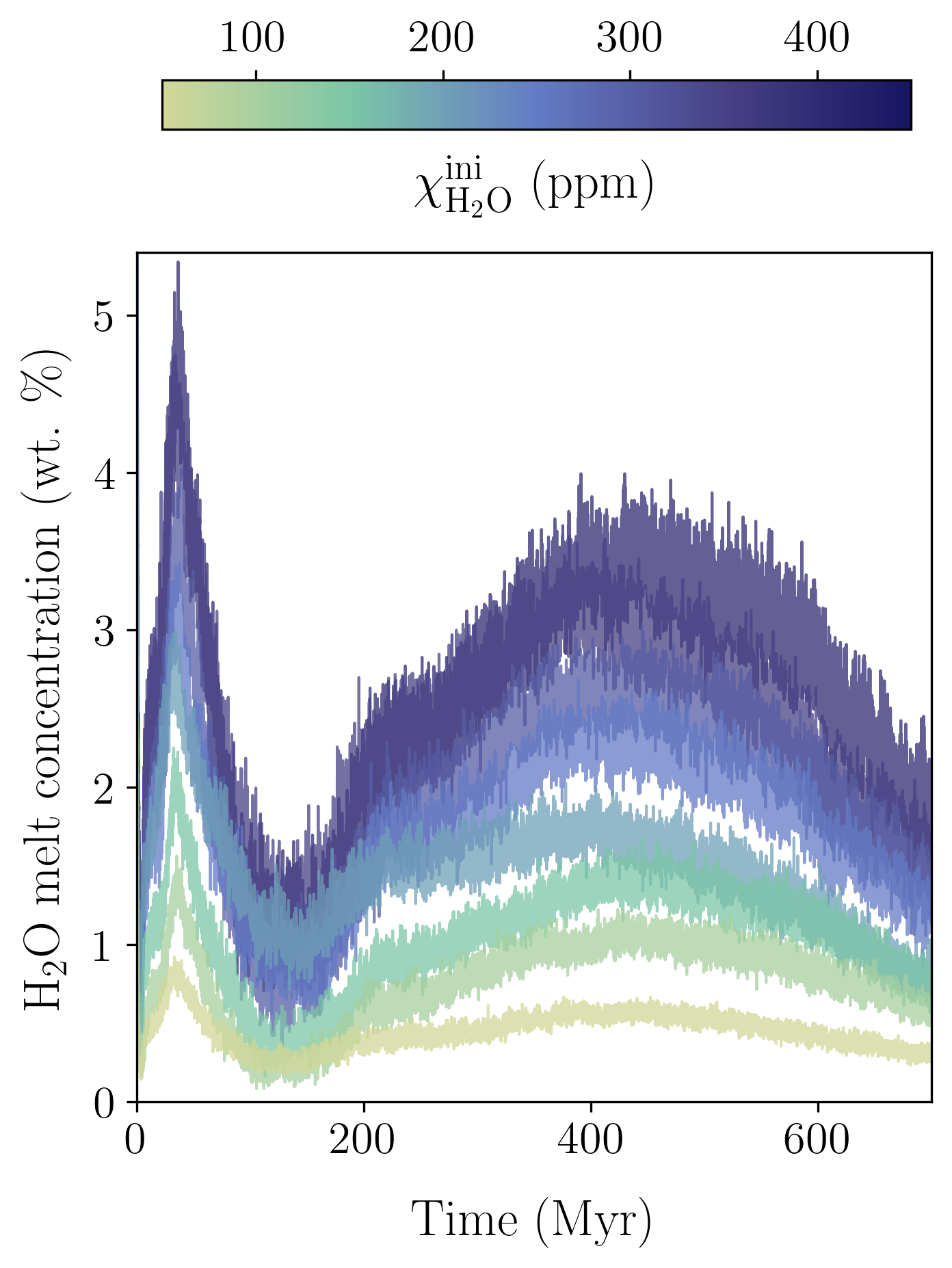} }}%
    \qquad
    \subfloat{{\includegraphics[width=0.465\linewidth]{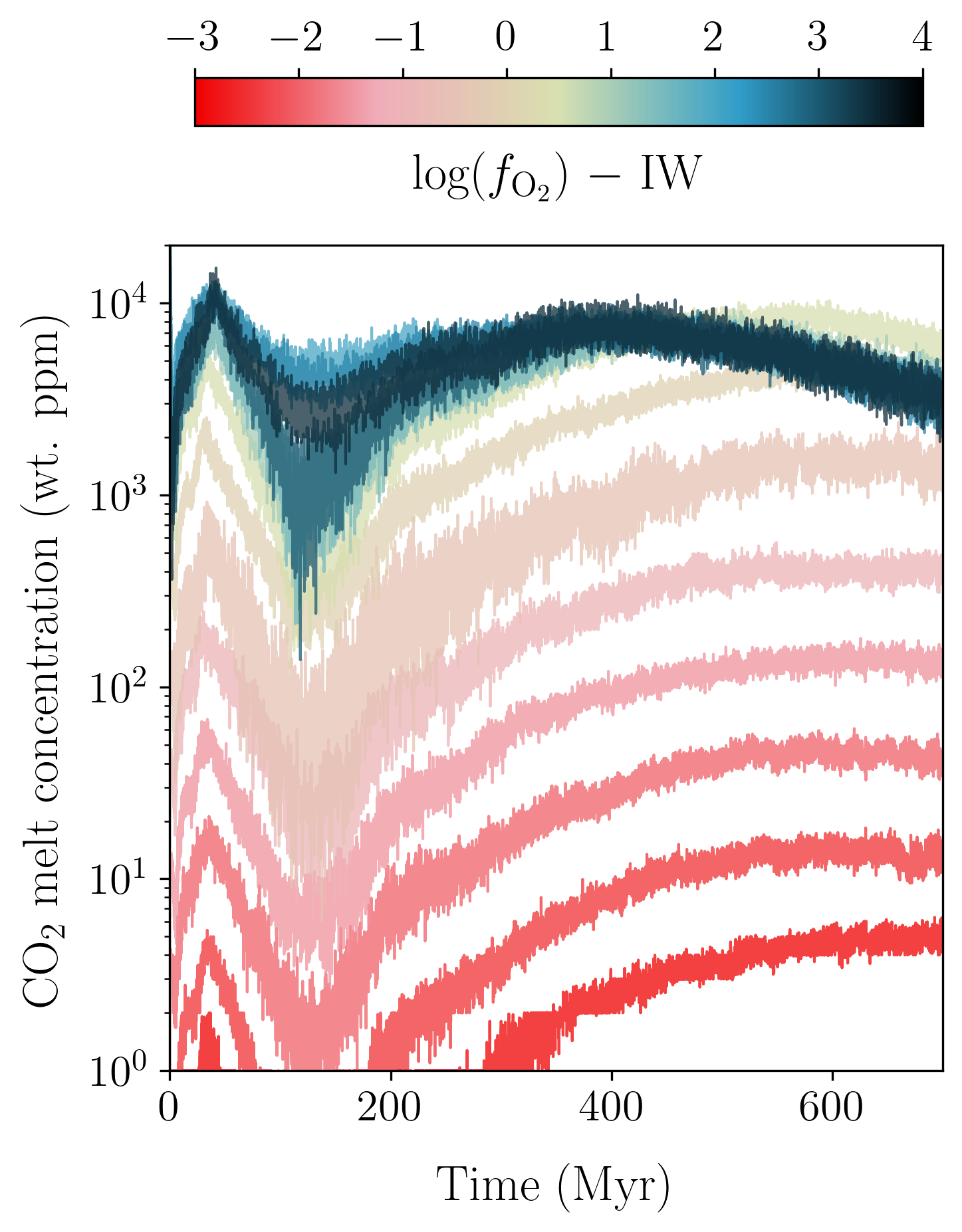} }}%
    \caption{The evolution of melt volatile contents for \ch{H2O} (\textit{left}; in weight percent) and \ch{CO2} (\textit{right}; in weight ppm). Cases are binned according to their values of the main input parameter controlling melt partitioning (see text): for \ch{H2O} this is the initial mantle \ch{H2O} content ($\chi_{\ch{H2O}}^{\rm{ini}}$), and for \ch{CO2} this is the mantle oxygen fugacity ($f_{\ch{O2}}$) with respect to the iron-w\"ustite buffer (IW). Lines are coloured by bin and show the mean melt concentration per time interval, given random values for the other input parameters.}%
    \label{fig:X_melt}%
\end{figure}

\begin{figure}
    \centering
    \includegraphics[width=1\linewidth]{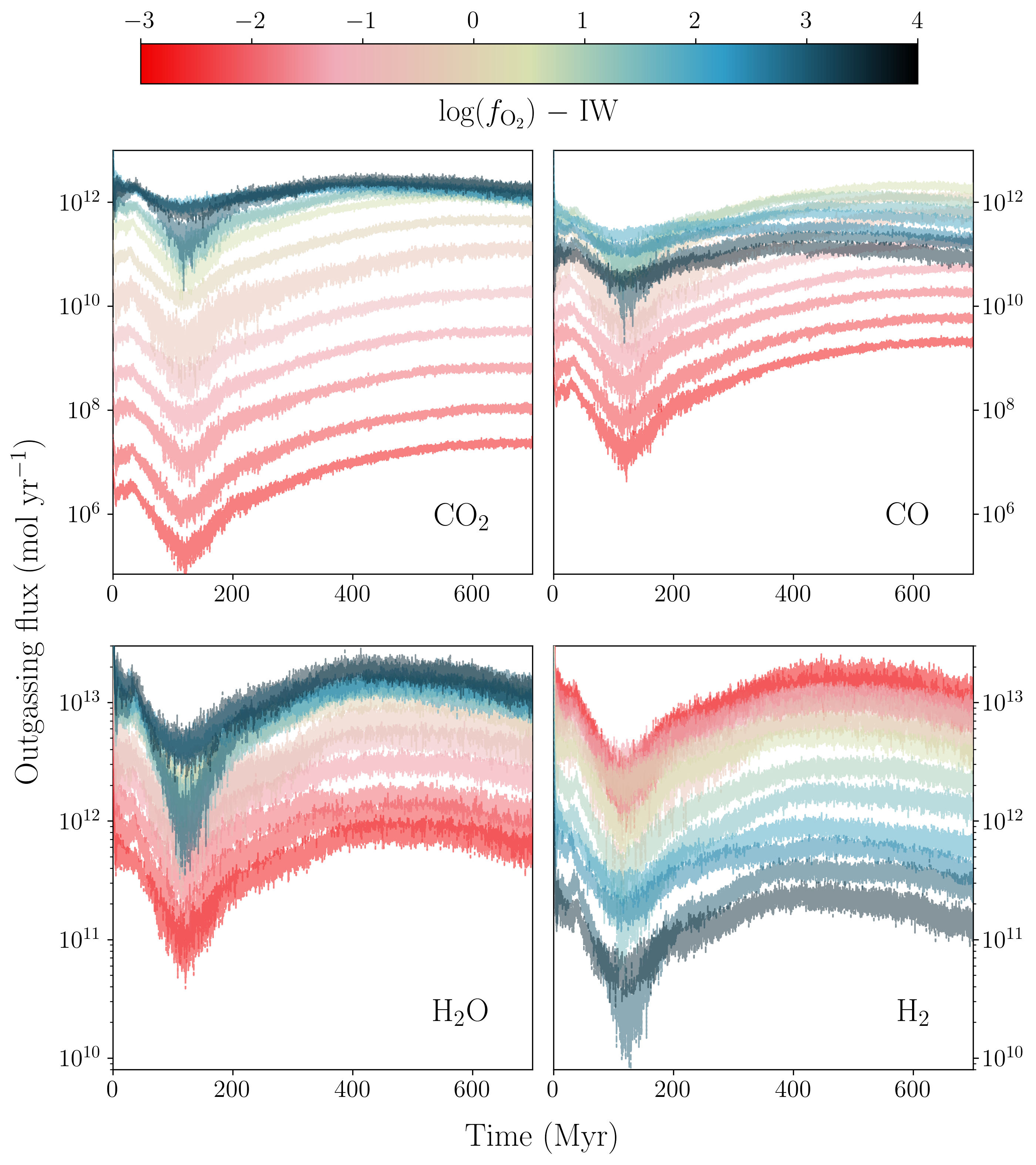}
    \caption{The mean evolution of volcanic outgassing fluxes for \ch{CO2} \textit{(top left)}, \ch{CO} \textit{(top right)}, \ch{H2O} \textit{(bottom left)}, and \ch{H2} \textit{(bottom right)}. Coloured lines from red to dark blue indicate increasingly oxidised mantles, binned to increments of 0.5 log-units in mantle oxygen fugacity ($f_{\ch{O2}}$) with respect to the iron-w\"ustite buffer (IW), and averaged across all other input parameters. }
    \label{fig:fO2_flux}
\end{figure}

Due to the different chemistry controlling the partitioning of C and \ch{H2O} into the melt (section \ref{sec:partitioning-methods}), $\chi_{\ch{H2O}}^{\rm melt}$ and $\chi_{\ch{CO2}}^{\rm melt}$ respond to different model parameters. These behaviours are reflected in figure \ref{fig:X_melt}. \ch{H2O} partitioning into the melt depends directly on the local mantle \ch{H2O} abundance, the maximum of which is set by $\chi_{\ch{H2O}}^{\rm ini}$. The effect of mantle content on melt content is non-linear: in equation (\ref{eq:h_partition}), $\chi_{\ch{H2O}}^{\rm melt}$ depends on the local melt fraction, whilst the presence of water facilitates melting by suppressing the solidus.

Carbon partitioning, in contrast, is strongly redox-dependent. This is true for the regime we model, where graphite is the stable phase, but would not apply to more oxidising conditions \citep{Stagno2019}. Note, that the local mantle source \ch{CO2} concentration can still limit $\chi_{\ch{CO2}}^{\rm melt}$. This effect is most obvious at very low values of $\chi_{\ch{CO2}}^{\rm ini}$ below 50~ppm, or for cases approaching maximum outgassing---in either situation the local carbon inventories can run out. \editthree{}{The most carbon-rich simulations can be seen (figure \ref{fig:X_melt}) to reach peak $\chi_{\ch{CO2}}^{\rm melt}$ before 700 Myr for this reason.} Meanwhile, the effect of $f_{\ch{O2}}$ on $\chi_{\ch{CO2}}^{\rm melt}$ is drastic everywhere. With each step of one log-unit below the IW buffer, we see $\chi_{\ch{CO2}}^{\rm melt}$ drop by an order of magnitude.

There is an early transient stage where the melt abundance drops steeply. This is associated with the early pulse of melting and hence depletion (figure \ref{fig:melt_vol}). 

\subsection{Exploration of outgassing scenarios}

A notable result of this work is that all scenarios, regardless of oxidation state, produce outgassing no higher than total estimates for modern Earth \citep[e.g.,][]{Catling2017}. To see why this is, here we examine what controls outgassing rates in this model.

\subsubsection{Outgassing as a function of mantle oxidation state}

As expected \citep{Holland1984, Kasting1993}, there is a heavy dependence of the outgassing rate on $f_{\ch{O2}}$. Figure \ref{fig:fO2_flux} shows the evolution of each species' mean outgassing flux in mol yr$^{-1}$, with cases binned by $f_{\ch{O2}}$ and all other input parameters varying according to a random uniform distribution (Table \ref{tab:inputs}). A more oxidised mantle is associated with more \ch{CO2} and \ch{H2O} outgassing, and a more reduced mantle is associated with more \ch{H2} outgassing. The pattern for \ch{CO} is more complicated. It peaks around IW: low $f_{\ch{O2}}$ limits the amount of total carbonate that can dissolve into the melt, despite the gas-phase equilibrium favouring \ch{CO} over \ch{CO2}. Outgassing rates increase slightly with time (by 10--15\%) during the 700 Myr modelled. The early pulse of outgassing is associated with the early pulse of melting (figure \ref{fig:melt_vol}).

\begin{figure}
    \centering
    \includegraphics[width=1\linewidth]{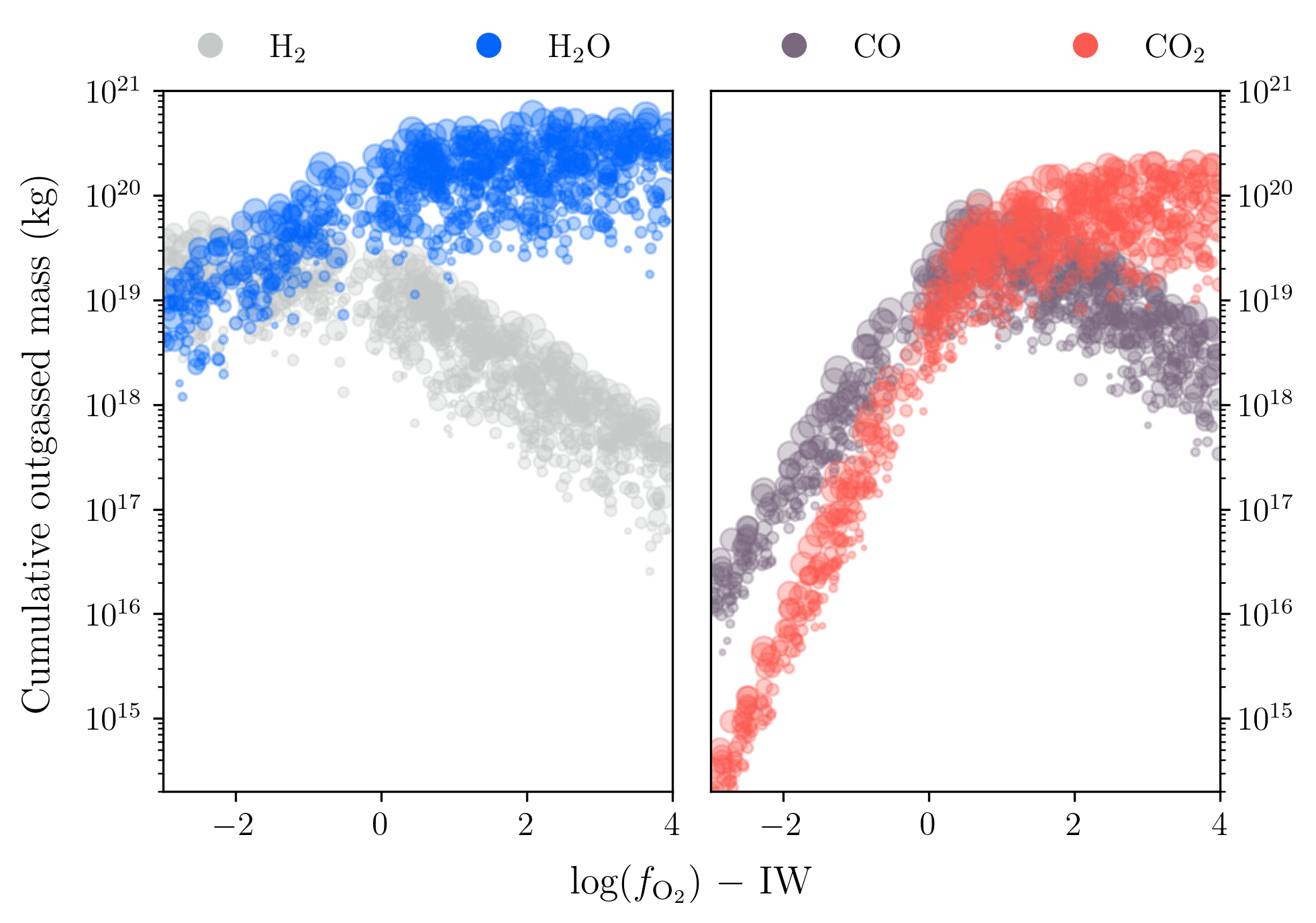}
    \caption{Each simulation's final cumulative outgassed masses of \ch{H2} (grey dots), \ch{H2O} (blue dots), \ch{CO} (aubergine dots), and \ch{CO2} (coral dots), as a function of the mantle oxygen fugacity ($f_{\ch{O2}}$) with respect to the iron-w\"ustite buffer (IW). Marker size increases with higher cumulative melt volume. All \Ncases~parameter combinations are shown.}
    \label{fig:fO2_scatter}
\end{figure}

To illustrate the dependence of evolved outgassing on $f_{\ch{O2}}$, figure \ref{fig:fO2_scatter} shows every run's cumulative outgassed mass per species, as a function $f_{\ch{O2}}$. Both C- and H-bearing species are well-separated in terms of mass, except for cases close to the IW buffer. Increasing $f_{\ch{O2}}$ from IW~$-$~3 to IW~$+$~4 is associated with an increase of over an order of magnitude in the mass of \ch{H2O}, and an increase of at least five orders of magnitude for \ch{CO2}, whilst the mass of \ch{H2} decreases. The hydrogen speciation shows a gradient in redox. The carbon speciation also shows a gradient in redox, whilst the total amount of carbon depends on partitioning (itself a function of redox).

It is worth emphasising that figure \ref{fig:fO2_scatter} does not show the actual gas masses residing in the atmosphere. For instance, the atmospheric partial pressure of \ch{H2O} is limited to the saturation vapour pressure of liquid water; on the Myr-timescales of this model, the cooling timescale of the outgassed plume to ambient atmospheric temperatures is irrelevant. At $T_{\rm surf} = 333$~K, the saturation vapour pressure is 0.2~bar. The cumulative mass of \ch{H2O} (which would quickly condense) is \editthree{less than}{at most about a tenth of} modern Earth's ocean mass, which would support the idea that earlier magma ocean degassing supplied a significant fraction of the original ocean mass \citep{Pahlevan2019}.

The influence of $f_{\ch{O2}}$ on the outgassed volatile masses is built into the model. Intrinsically, there are three reasons for the mass dependence in figure \ref{fig:fO2_scatter}. The first two reasons comprise the redox-dependent speciation (setting total carbon) and volatile speciation (setting [\ch{H2O}]/[\ch{H2}] and [\ch{CO2}]/[\ch{CO]}) already described. The third reason is that the reduced gases investigated here have a lower molecular mass than their oxidised counterparts. Thus, for equal quantities outgassed, reduced atmospheres are necessarily less massive than oxidised atmospheres.

\subsubsection{Other factors influencing outgassing}

\begin{table*}
\caption{Final partial pressures, masses, and fluxes for each species after 700~Myr of outgassing, binned by mantle oxygen fugacity ($f_{\ch{O2}}$) with respect to the iron-w\"ustite buffer (IW). Fluxes represent averages over the last 10~Myr. Results are shown as the bin medians, with 1$\sigma$ limits super- and subscripted.}       
\label{tab:results}     
\begin{adjustbox}{width=1\textwidth}
\centering                         


\begin{tabular}{>{\centering\arraybackslash}m{3cm} l p{0mm} *{7}{r}}
\hline\hline
\noalign{\vskip 1mm}
 & & & \multicolumn{7}{c}{log($f_{\mathrm{O}_2}$) $-$ IW} \\
\cmidrule(lr){4 - 10}
& & & [-3, -2)& [-2, -1)& [-1, 0)& [0, 1)& [1, 2)& [2, 3)& [3, 4] \\
\noalign{\vskip 1mm} \hline \noalign{\vskip 1mm}
\textsc{Final pressure}
 & 
CO$_2$ & 
& $0.0_{-0.0}^{+0.0}$
& $0.0_{-0.0}^{+0.0}$
& $0.0_{-0.0}^{+0.1}$
& $1.2_{-0.8}^{+1.4}$
& $3.5_{-1.8}^{+3.1}$
& $5.9_{-3.3}^{+6.3}$
& $7.2_{-4.5}^{+6.8}$
\\
bar
 & 
CO & 
& $0.0_{-0.0}^{+0.0}$
& $0.0_{-0.0}^{+0.0}$
& $0.1_{-0.1}^{+0.3}$
& $2.1_{-1.3}^{+2.3}$
& $2.4_{-1.4}^{+2.3}$
& $1.2_{-0.6}^{+1.8}$
& $0.5_{-0.3}^{+0.6}$
\\
 & 
H$_2$O & 
& $0.3_{-0.2}^{+0.3}$
& $1.3_{-0.7}^{+1.6}$
& $5.1_{-2.7}^{+7.0}$
& $21.3_{-11.7}^{+18.7}$
& $32.2_{-17.0}^{+26.2}$
& $38.5_{-20.3}^{+34.0}$
& $51.0_{-30.6}^{+31.4}$
\\
 & 
H$_2$ & 
& $5.7_{-3.5}^{+4.2}$
& $6.4_{-3.0}^{+7.1}$
& $8.9_{-3.8}^{+10.0}$
& $10.0_{-5.2}^{+7.9}$
& $5.2_{-2.8}^{+3.7}$
& $2.0_{-1.1}^{+2.0}$
& $0.8_{-0.5}^{+0.6}$
\\
\noalign{\vskip 1mm} \hline \noalign{\vskip 1mm}
\textsc{Final mass}
 & 
CO$_2$ & 
& $0.0_{-0.0}^{+0.0}$
& $0.0_{-0.0}^{+0.1}$
& $0.5_{-0.3}^{+1.3}$
& $15.3_{-9.6}^{+19.0}$
& $40.7_{-21.5}^{+30.2}$
& $59.6_{-29.9}^{+52.0}$
& $66.7_{-38.2}^{+72.1}$
\\
kg ($\times 10^{18}$)
 & 
CO & 
& $0.0_{-0.0}^{+0.0}$
& $0.3_{-0.2}^{+0.5}$
& $2.2_{-1.2}^{+3.9}$
& $19.0_{-10.9}^{+20.6}$
& $16.1_{-8.3}^{+15.7}$
& $7.5_{-3.5}^{+10.0}$
& $3.1_{-1.9}^{+3.8}$
\\
 & 
H$_2$O & 
& $9.4_{-5.5}^{+7.8}$
& $23.9_{-11.9}^{+27.1}$
& $57.9_{-27.1}^{+60.4}$
& $128.7_{-74.1}^{+106.6}$
& $154.9_{-93.6}^{+129.0}$
& $171.7_{-102.4}^{+153.2}$
& $220.2_{-142.1}^{+137.2}$
\\
 & 
H$_2$ & 
& $18.2_{-10.4}^{+18.2}$
& $13.3_{-6.2}^{+14.3}$
& $12.2_{-6.2}^{+10.5}$
& $6.8_{-3.7}^{+5.7}$
& $2.8_{-1.7}^{+2.3}$
& $1.0_{-0.6}^{+0.9}$
& $0.4_{-0.2}^{+0.3}$
\\
\noalign{\vskip 1mm} \hline \noalign{\vskip 1mm}
\textsc{Final flux}
 & 
CO$_2$ & 
& $0.0_{-0.0}^{+0.0}$
& $0.0_{-0.0}^{+0.0}$
& $0.0_{-0.0}^{+0.1}$
& $0.7_{-0.4}^{+1.2}$
& $1.3_{-0.8}^{+1.4}$
& $1.4_{-0.8}^{+1.6}$
& $1.8_{-1.2}^{+2.3}$
\\
mol yr$^{-1}$ ($\times 10^{12}$)
 & 
CO & 
& $0.0_{-0.0}^{+0.0}$
& $0.0_{-0.0}^{+0.0}$
& $0.2_{-0.1}^{+0.4}$
& $1.5_{-0.8}^{+1.8}$
& $0.9_{-0.5}^{+1.1}$
& $0.3_{-0.2}^{+0.5}$
& $0.1_{-0.1}^{+0.2}$
\\
 & 
H$_2$O & 
& $0.8_{-0.6}^{+0.9}$
& $1.9_{-1.2}^{+1.8}$
& $4.6_{-2.6}^{+4.2}$
& $9.1_{-5.4}^{+8.6}$
& $11.4_{-6.5}^{+13.1}$
& $11.3_{-7.8}^{+10.7}$
& $14.3_{-7.4}^{+15.9}$
\\
 & 
H$_2$ & 
& $12.9_{-8.1}^{+15.0}$
& $8.7_{-4.8}^{+7.6}$
& $8.2_{-5.0}^{+7.4}$
& $4.3_{-2.6}^{+3.7}$
& $1.8_{-1.1}^{+2.4}$
& $0.5_{-0.3}^{+0.8}$
& $0.2_{-0.1}^{+0.3}$
\\
\noalign{\vskip 1mm}
\hline
\end{tabular}


\end{adjustbox}
\end{table*}

\begin{figure}
\centering
  \includegraphics[width=0.9\linewidth]{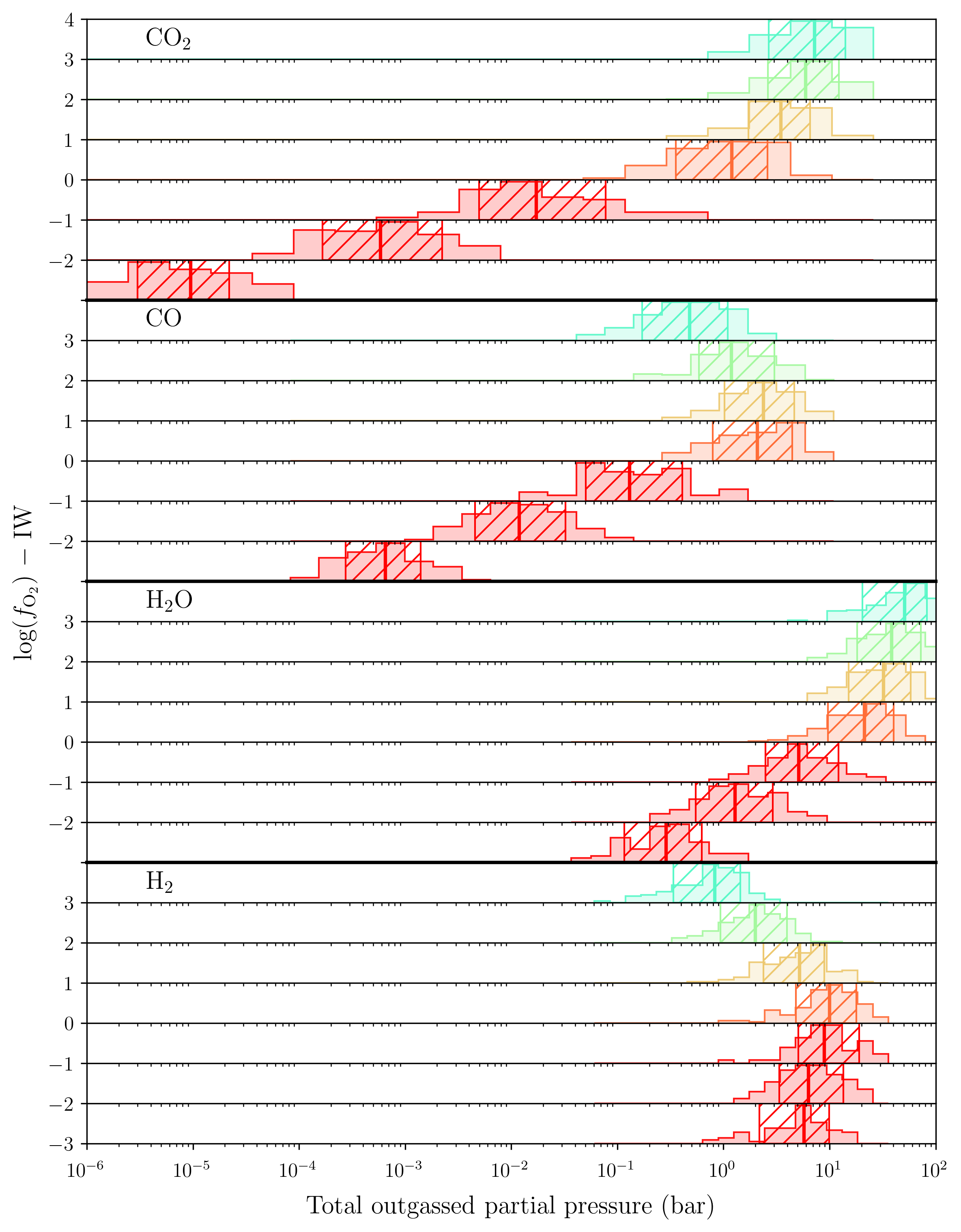}
\caption{Histograms of the empirical distribution of 700-Myr-cumulative outgassed masses for (from top to bottom) \ch{CO2}, \ch{CO}, \ch{H2O}, and \ch{H2}. Distributions are marginalised across the mantle oxygen fugacity ($f_{\ch{O2}}$) with respect to the iron-w\"ustite buffer (IW), where each colour shows one of five log($f_{\ch{O2}}$) bins as indicated by the $y$-axis ticks. Bold vertical lines indicate the medians; hatched regions mark the 1$\sigma$ width.\label{fig:hist_mass_fo2}}
\end{figure}

\begin{figure}
\centering
  \includegraphics[width=\linewidth]{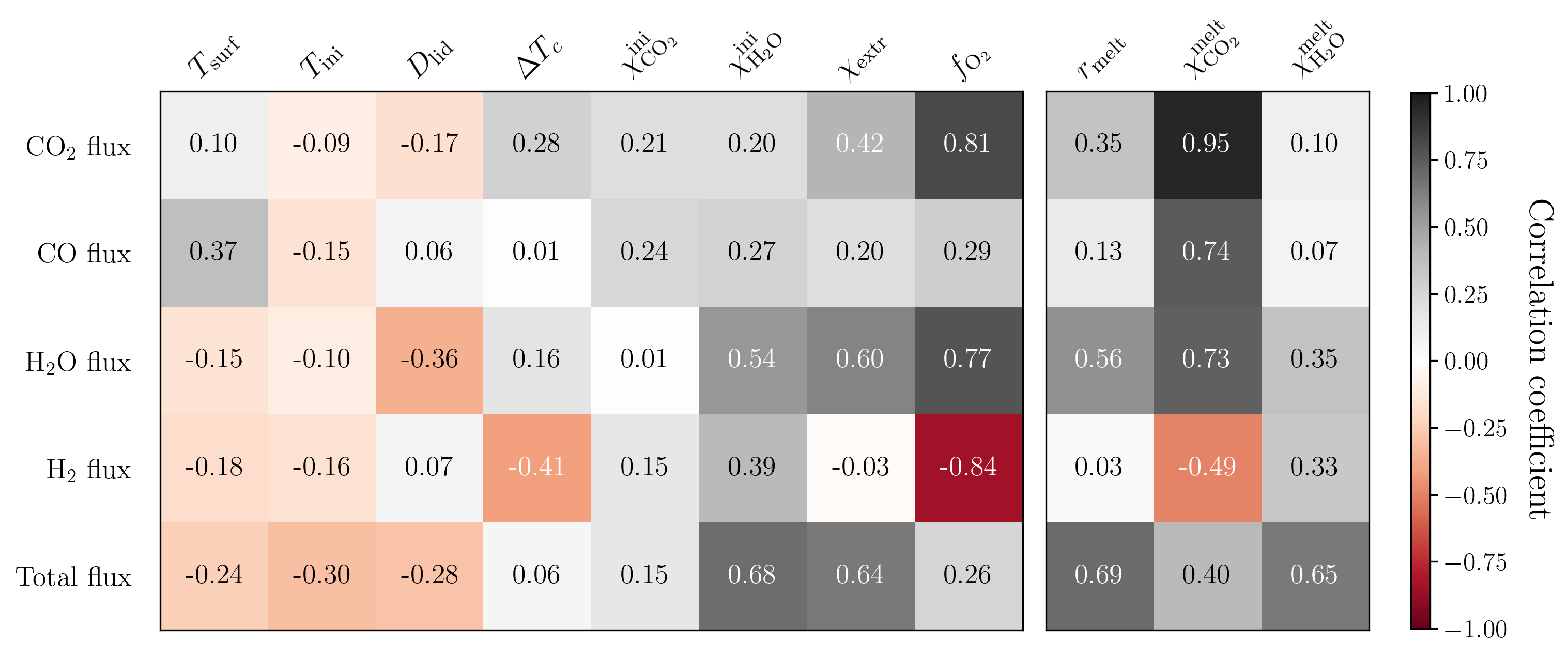}
\caption{\edittwo{}{\textit{(Left):}} The matrix of Spearman's rank correlation coefficients between input parameters and outgassing fluxes averaged over the final 10~Myr, plus the summed flux of all outgassed molecules for the same time frame. Symbols are defined in Table \ref{tab:inputs}. \edittwo{}{\textit{(Right):} The same, but for key intermediate output variables---the volumetric melt production rate, $r_{\rm melt}$, and the melt concentrations of \ch{CO2} and \ch{H2O}, $\chi_{\ch{CO2}}^{\rm melt}$ and $\chi_{\ch{H2O}}^{\rm melt}$. Note that the moderate correlation between $\chi_{\ch{CO2}}^{\rm melt}$ and H-species outgassing is due to the mutual effect of $f_{\ch{O2}}$ on both quantities, while the correlation of $f_{\ch{O2}}$ with the total flux appears low because it does not affect the sum of \ch{H2} and \ch{H2O}.}\label{fig:corr}}
\end{figure}

Figure \ref{fig:corr} summarises the Spearman's rank (nonlinear) correlation coefficients between each species' outgassing flux, averaged over the final 10 Myr, and the eight input parameters plus the melt volumes and volatile concentrations. \editthree{Beyond the mantle oxidation state, we find relatively moderate correlation with $\chi_{\ch{H2O}}^{\rm ini}$ and $\chi_{\rm extr}$, weak correlation with $\chi_{\ch{CO2}}^{\rm ini}$, and no correlation with the initial thermal state.}{This section explains how the remaining correlations stem from the model.}

\paragraph{Influence of initial temperatures}

We find virtually no correlation between initial temperature conditions and outgassing fluxes after 700~Myr of convection. However, an initially hotter mantle closer to the solidus temperature forces a sharp pulse of melting in the first 100 Myr, expelling massive amounts of water and indeed raising the total mass of water outgassed. This is also not to say that instantaneous local temperatures are unimportant, as they affect the melt fraction as well as the equilibrium constants in (\ref{eq:c_partition}), (\ref{eq:equilibrium_constant_hydrogen}), and (\ref{eq:equilibrium_constant_carbon}).

\paragraph{Influence of extrusive volcanism fractions}

From (\ref{eq:m_atm}), we expect an approximately linear relationship between the maximum outgassing flux and the fraction of melt allowed to contribute to outgassing. Although the correlation between this fraction and the computed outgassing fluxes is not as strong as $f_{\ch{O2}}$, accounting for extrusive volcanism in this model automatically downscales outgassing. For example, all else being equal, the difference between $\chi_{\rm extr}=10$\% and $\chi_{\rm extr}=100$\% would be an order of magnitude in outgassing.

\paragraph{Role of melt production rate versus melt volatile content}

Figure \ref{fig:corr} also shows the correlations of all outgassing fluxes with the instantaneous volatile melt concentrations and melt production rates. \ch{CO2} outgassing fluxes correlate more strongly with \ch{CO2} melt concentrations than with the total melt production rates, whilst the opposite appears to be true for \ch{H2O}. Note that \ch{H2O} melt concentrations are not independent of melt volumes because of partitioning of water into the melt---\ch{H2O} concentrations are largest for the smallest melt fractions. \editthree{}{To some degree, the relative strengths of the correlations reflect the experimental variances in melt production and volatile melt contents, given the ranges of the prior distributions on the input parameters. Nevertheless, in our simulations, outgassing will be ultimately limited by the volatile budget of the mantle. Thus rates of outgassing would not scale indefinitely with rates of volcanism.}

\editthree{Therefore, simply scaling outgassing with melting may neglect the high variability of volatile content in melts and could misrepresent fluxes in a stagnant lid regime. Whilst this is especially true for carbon species, where melt content overwhelmingly depends on $f_{\ch{O2}}$, it also holds for \ch{H2O} and \ch{H2}, and so is not restricted to redox-dependent partitioning.}{}

\subsubsection{\editthree{Composition, flux, and total mass of early Archean outgassing}{Redox-marginalised outgassing distributions}}

Figure \ref{fig:hist_mass_fo2} presents the experimental distributions of each species' cumulative outgassed partial pressures, shown for different redox bins. Again, the dominant effect of $f_{\ch{O2}}$ is visible in how these distributions are separated along the $x$-axis. The median and 1$\sigma$ values of these partial pressures are quantified in Table \ref{tab:results}, along with the corresponding masses and fluxes.

\subsection{Secular mantle oxidation}

\begin{figure}
\centering\includegraphics[width=1\linewidth]{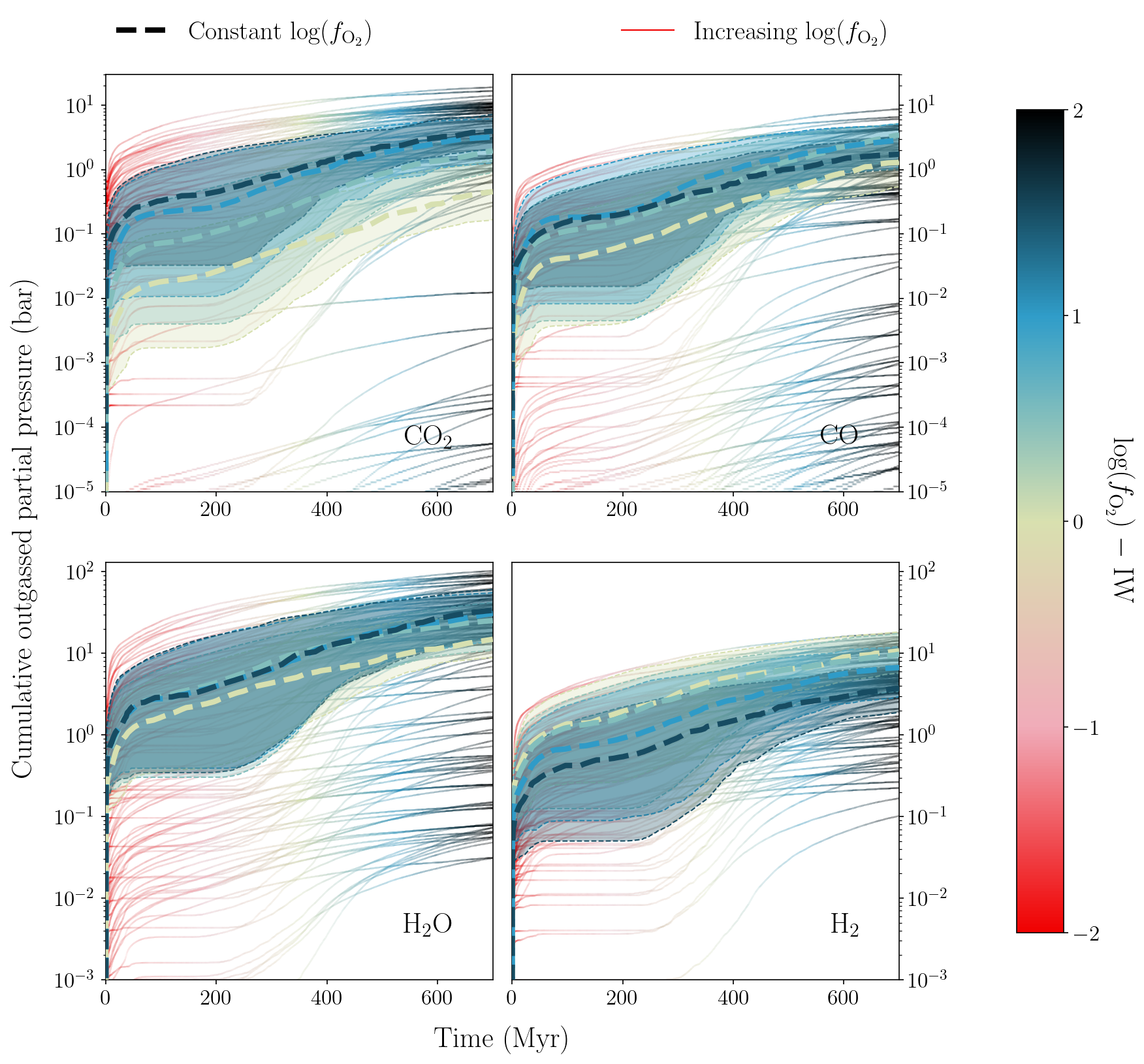}
\caption{The evolution of the cumulative outgassed partial pressure for scenarios where the mantle oxygen fugacity ($f_{\ch{O2}}$) with respect to the iron-w\"ustite buffer (IW) increases log-linearly from IW~$-$~2 to IW~$+$~2 over 700~Myr (thin solid gradient lines; $N=110$). For comparison, also shown are the $f_{\ch{O2}}$-bin medians of previous scenarios where $f_{\ch{O2}}$ is fixed randomly between IW and IW~$+$~2 (thick dashed lines; $N=147$), where the thin dashed lines spanned by swaths indicate 1$\sigma$ above and below the median. Subplots show the species \ch{CO2} \textit{(top left)}, \ch{CO} \textit{(top right)}, \ch{H2O} \textit{(bottom left)}, and \ch{H2} \textit{(bottom right)}. The colour indicates the instantaneous mantle redox. \label{fig:evol_fo2increase}}
\end{figure}

Several mechanisms for mantle oxidation have been proposed \citep[e.g.,][]{Wood2006, Sharp2013, Gaillard2014, Wordsworth2018,  Schaefer2018, NICKLAS2019}, including degassing itself \citep{Kasting1993}. These mechanisms would be associated with different durations. By tracking the concentration of abundant multivalent cations, one might self-consistently evolve the mantle redox state within our outgassing scenarios. We leave this to future work, and as a first step, simply consider a linear increase of $\log (f_{\ch{O2}})$ from IW~$-$~2 to IW~$+$~2 over 700~Myr, for a new set of 110 runs with otherwise random parameters as before (Table \ref{tab:inputs}). 

Figure \ref{fig:evol_fo2increase} demonstrates that secularly increasing $\log (f_{\ch{O2}})$ results in cumulative outgassing not obviously distinguishable from a constant $\log (f_{\ch{O2}}) - {\rm IW} \in [0, 2]$. The spread in these cases would be largely explained by variations in the melting rate.

Low $f_{\ch{O2}}$ does not affect melt concentrations of H-species, such that the same total amount of total H can be outgassed. Carbon, meanwhile, is strongly limited in melts during the early reduced stage, which might imply lower cumulative partial pressures. However, Figure \ref{fig:evol_fo2increase} suggests that this effect is somewhat muted due to the scatter induced by other unknown parameters. Note, regardless, that small differences in cumulative outgassing could represent dramatic differences in the ultimate atmospheric composition; scenarios of evolving mantle redox likely deserve more rigorous treatment than attempted here.


\section{Discussion}

\subsection{Some important considerations about the assumptions in this model}

\subsubsection{Viscosity treatment}

\editthree{}{This work has considered a single set of upper mantle rheological parameters, corresponding to the canonical Arrhenius viscosity law for wet olivine from \citet{Karato1993}. Lower mantle rheology is taken from \citet{Tackley2013}.  Different viscosity treatments could potentially influence melting rates and therefore outgassing rates.} 

\editthree{}{However, several regulation mechanisms exist such that an immediate weakening of upper mantle viscosity, whilst increasing the convective velocity, may not lead to dramatically higher melting in the long term. Firstly, latent heat loss associated with any melting that does occur would cool the mantle (the latent heat of melting is indeed modelled here). These lowered temperatures would both limit further melting and stiffen the local viscosity in a negative feedback loop \citep{Ogawa2011}. Secondly, a sudden increase in convective velocity would lead to faster replenishment of depleted material in the upper mantle, which promotes melting; yet, thirdly, this effect would be offset simultaneously by more efficient cooling of the mantle, which suppresses melting.}

\editthree{}{The \citet{Karato1993} law for wet diffusion creep does have a relatively weak temperature dependence and a strong pressure dependence, which could reduce some of the feedbacks described here. Nevertheless, as in \citet{dorn2018outgassing}, we expect that melting is ultimately controlled by the lower mantle viscosity because rapid flow in the upper mantle merely leads to rapid depletion. With that said, we would still emphasise that the results here are subject to our chosen viscosity treatment, and should be interpreted as such.}

\subsubsection{Submarine versus subaerial outgassing}

We have assumed that all outgassing is subaerial and occurs at 1~bar. Several independent constraints place an upper limit on either the surface barometric pressure or $p\ch{N2}$ at 1.1~bar for 3.5--2.7~Ga \citep{Marty2013, Catling2020}. A lack of plate motion might allow volcanic eruptions to construct topography above sea level relatively quickly (e.g., Olympus Mons, a Martian shield volcano comparable to Germany in area). In this case, volatiles would degas freely from surface lavas at atmospheric pressure. Because near 1~bar there is almost no effect on gas solubility or $f_{\ch{O2}}$ in our chosen parameterisations, we do not consider solubility or any redox change during degassing itself. If most outgassing occurred on the seafloor, then the higher pressures would allow melts to retain more volatiles, as long as the gas and melt are in equilibrium (e.g., eruptions are not explosive). Global outgassing rates would then be lower than predicted here, and show a different ratio of outgassing species \citep{Gaillard2011}.

\subsubsection{Note on \ch{CH4} outgassing}

We neglect the outgassing of \ch{CH4} entirely because its vapour phase is not stable at the pressure, temperature, and redox ranges we consider \citep{zhang2009model, wetzel2013degassing, ramirez2014warming}. Some \ch{CH4} fluid could be produced from a reduced source at high pressures up to 11~GPa \citep{Scott2004}. Although \ch{CH4} can also be stable at surface pressures, melt temperatures would be too high for it to degas directly.

\subsubsection{Buoyancy limit of melting}

\editthree{}{As in \citet{noack2014can, Noack2017}, we have not considered the possibility of melting at depths greater than 12~GPa. This assumes that basaltic melt is not buoyant above this pressure because it is denser than olivine \citep{Ohtani1995}. However, \citet{Mosenfelder2009} have suggested that basaltic melt can nevertheless remain less dense than ringwoodite and wadsleite up to $\sim$25 GPa. Therefore the actual density crossover pressure could be lower than modelled here \citep[e.g.,][]{Beuchert2013}. This could increase the volume of the mantle annulus in which melting can occur, possibly leading to enhanced outgassing if all buoyant melt at depth reaches the base of the lithosphere. However, it may be that no additional melting occurs at these depths, depending on how the local mantle temperatures compare to the (strongly pressure-dependent) melting temperature.}

\subsubsection{Stagnant lid outgassing effects}\label{sec:stagnant-lid}

Some general consequences of imposing a stagnant lid regime are listed below. This compilation is by no means complete and would benefit from future work. \citet{noack2014can}, \citet{tosi2017habitability}, \citet{Foley2018}, and \citet{Gaillard2021} also discuss how tectonic regimes can affect outgassing rates. The arguments here could be tentatively supported by the observation that Venus currently appears less volcanically-active than Earth \citep{Smrekar2010}. 

\begin{enumerate}[i.]

\item \textit{Interior temperatures:} For a constant mass, stagnant lid planets have hotter mantles because they lack the efficient cooling by subducting plates. This would enhance melting and volcanism.

\item \textit{Melt volumes:} However, stagnant lid planets may be associated with less continuous melting \textit{for the same interior temperature} because, three-fold, the melt zone is separated from the surface by a thick lithosphere, convection is weaker due to the lower temperature contrast, and the depleted mantle is not refertilised by subducting plates.

\item \textit{Volatile inventories of the magma source:} Higher mantle volatile contents would lead to higher melt contents insofar as they reflect each other. The \editthree{magmas generated}{mantle source} supplying stagnant lid volcanism \editthree{dissolve less \ch{CO2} and \ch{H2O}}{should be less volatile-rich} than \editthree{those}{the source} of arc volcanism, which contains subducted crustal material relatively heavy in carbonates, organic carbon, and water \citep{WALLACE2005}. The mid-ocean ridge source is drier, yet still generally appears carbon-rich compared to the initial concentrations we have adopted for the early degassed, differentiated mantle \citep{Hauri2019}. Hotspots, meanwhile, sample the deep mantle, which seems to contain both lingering nebular and recycled components \citep[e.g.,][]{MILLER2019}. Analyses of ocean island basalts suggest even higher mantle source concentrations than mid-ocean ridges---likewise associated with high-\ch{CO2} melts \citep[e.g.,][]{SHORTTLE2015, TUCKER2019, Broadley2019, MILLER2019}\editthree{}{, as are ridge segments in proximity to hotspots \citep{Voyer2019}}.  

\item \textit{Upper mantle hydration:} If subducted oceanic crust is wet, then the presence of water would depress the solidus \citep{Katz2003}. This could facilitate melting near subduction zones in a plate tectonics regime.
\end{enumerate}

\subsection{Observational context}

\begin{figure}
\captionsetup[subfigure]{labelformat=empty}
\centering
    \subfloat[]{\label{sublable1}\includegraphics[width=1\linewidth]{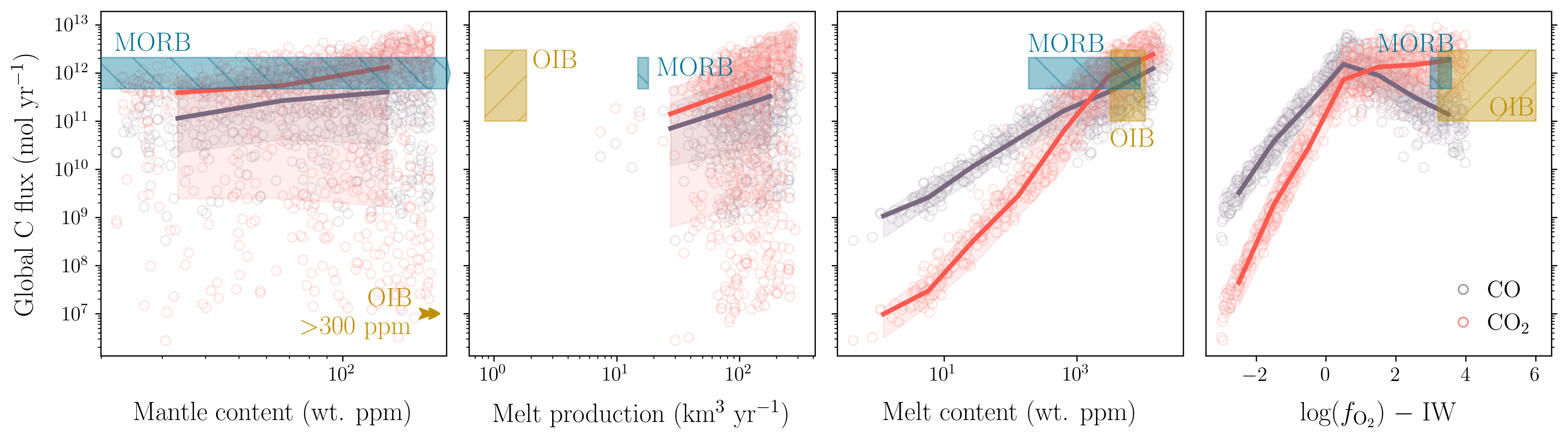}} \\
\caption{\label{fig:summary_C}
Summary of outgassing fluxes of CO (blue) and CO$_2$ (red), with respect to (from left to right) mantle source CO$_2$ content, melt production rate, melt CO$_2$ content, and mantle oxygen fugacity ($f_{\ch{O2}}$) relative to the iron-w\"ustite (IW) buffer. The mantle source concentrations in our model refer to the maximum (initial) values. All variables save $f_{\ch{O2}}$ are represented by the final 10-Myr mean. Each hollow circle denotes an individual model run. Solid lines show the median of all runs, and swaths show the 1$\sigma$ deviation. For context, \editthree{modern Earth estimates are also shown for \ch{CO2}. The light blue rectangle spans the estimate from \citet{Hauri2017} of the CO$_2$ outgassing flux from the global mid-ocean ridge (MOR) system. The beige rectangle spans the same estimate from \citet{DASGUPTA2010} for hotspot volcanism. MOR mantle source concentrations are from \citet{Hauri2017}; depleted mantle (DM) concentrations are from \citet{Marty2012}; ocean island basalt (OIB) mantle source concentrations are from \citet{Hauri2019}; magma supply rates are from \citet{Mjelde2010} and \citet{Cogne2006} for 19 hotspots and the global MOR system respectively; melt concentrations of CO$_2$ are from \citet{Hauri2019}; and source}{we include estimates of modern Earth's CO$_2$ outgassing, spanned by blue rectangles for the mid-ocean ridge (MOR) system, and by beige rectangles for hotspots. MOR estimates are taken from the data in \citet{Voyer2019} and \citet{Hauri2019}, where the ranges of outgassing fluxes and melt production rates correspond to their quoted uncertainty on the global total, and the ranges of mantle and melt contents correspond to the 2$\sigma$-width of their log-normal sample distribution. For hotspots, the \ch{CO2} flux lower limit is the sum from major hotspots from \citet{Hauri2019}, and the generous upper limit is taken from \citet{Marty1998}; ocean island basalt (OIB) mantle source and melt concentrations are from \citet{Hauri2019}; magma supply rates are the sum over 19 hotspots from \citet{Mjelde2010}. Estimates of the mantle source} $f_{\rm O_2}$ are from \citet{ONEILL2018} and \citet{AMUNDSEN1992} for MORB and OIB respectively. 
}
\end{figure}

\begin{figure}
\centering\includegraphics[width=1\linewidth]{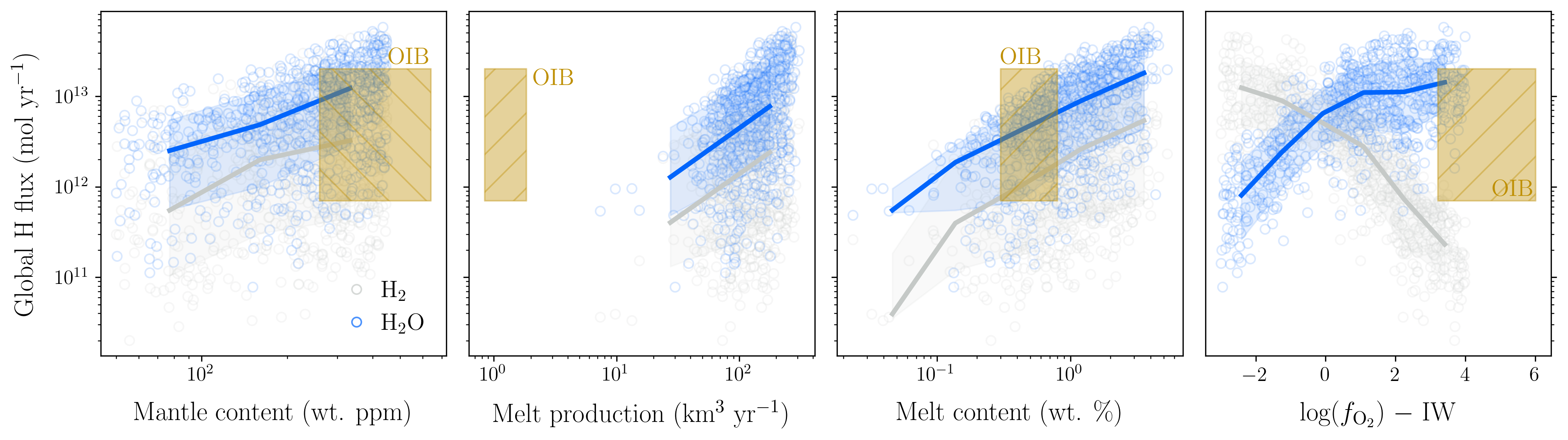}
\caption{\label{fig:summary_H}Analogous to Figure \ref{fig:summary_C}, but for H-bearing species, with H$_2$ flux in grey and H$_2$O flux in blue. The beige rectangle spans the estimate from \citet{DASGUPTA2010} for CO$_2$ hotspot volcanism, multiplied by 7--8, the typical ratio of H$_2$O to CO$_2$ in hotspot volcanic gas from \citet{Holland1984}; i.e., following \citet{Catling2017}. Mantle source concentrations refer to Hawaii estimates from \citet{Wallace1998}; melt concentrations are also Hawaii estimates from \citet{WallaceAnderson1998}. Mid-ocean ridge volcanism is omitted because it is known to be very dry, compared to the range shown here. Other data are explained in the caption to figure \ref{fig:summary_C}.}
\end{figure}

\subsubsection{Modern Earth's volcanic outgassing}

\editthree{It might help bring context to the results here if we imagine how they collate with modern volcanic outgassing. This would require comparing the same ends across different means, however; modern outgassing is almost entirely inseparable from plate tectonics (see section 4.3 above). Hence, a comparison would mostly provide a platform to understand the trade-offs between different components of the outgassing flux.}{}

\editthree{}{A key result from this work is that our stagnant lid outgassing model produces average outgassing rates no higher than those estimated for modern Earth under plate tectonics. Whereas $\sim$68\% of our cases more oxidised than IW + 3 have a \ch{CO2} flux of between 0.6 and 4.1 Tmol~yr$^{-1}$, estimates for Earth's global \ch{CO2} outgassing now rest at around $8.5 \pm 2$~Tmol~yr$^{-1}$ \citep{Catling2017}. Metamorphic degassing and arc volcanism, irrelevant to a stagnant lid regime, both account for a quarter of this global estimate. Although our model does not necessarily apply to hotspot or mid-ocean ridge outgassing in a plate tectonics regime either, estimates of the modern rates of these processes might provide a useful point of context for illustrating the mechanisms that keep our outgassing results relatively low.}

\editthree{}{Namely, we see four main factors potentially influencing outgassing: mantle volatile content, melt production rates, melt volatile contents, and mantle oxidation state.} To this end, figures \ref{fig:summary_C} and \ref{fig:summary_H} summarise our modelled outgassing rates as a function of these four factors, overplotted by the relevant present-day estimates for hotspot and ridge volcanism.

\editthree{}{It should be noted upfront that our model may not be able to correctly capture modern ridge and hotspot outgassing rates, especially for \ch{CO2}. As explained in section \ref{sec:stagnant-lid} above, these processes sample recycled material generally having both (i) higher oxidation states, at the QFM buffer or above \citep{AMUNDSEN1992, MOUSSALLAM2016, ONEILL2018, MOUSSALLAM2019}, to which our carbon partitioning model does not apply, and (ii) higher source volatile concentrations than our quasi-primordial bulk mantle values \citep{Hauri2019}. The combined effect of (i) and (ii) is that \ch{CO2} can potentially reach much higher melt concentrations on average. In oxidised conditions where carbonate is stable, abundant carbon would be lost very efficiently during partial melting.}

\editthree{}{Further to this point, modern outgassing rates are quite difficult to measure. Most quoted outgassing rates come from multiplying estimates of the melting rate with estimates of the concentration of volatiles in the melt. The volatile contents of melts are largely uncertain because melts are partially degassed; estimates rely on the behaviour of geochemical proxies \citep[see][]{Michael2015}. An in-depth review of Earth's ridge and hotspot outgassing rates is well beyond the scope of this work.}

\editthree{}{\paragraph{Mantle oxidation state} Whilst shown here for completeness, mantle $f_{\ch{O2}}$ does not play an important role in controlling modern \ch{CO2} or \ch{H2O} outgassing because the modern upper mantle is oxidised essentially everywhere, such that \ch{H2} and \ch{CO} would never be favoured, and carbon partitioning in the melt is not redox-dependent.} 

\editthree{}{\paragraph{Melt production} Estimates of globally-integrated magma supply rates at mid-ocean ridges tend to show better agreement---most not far from 20 km$^3$ yr$^{-1}$ \citep{Crisp1984, Mjelde2010, Voyer2019}---although these estimates refer to the volume of all melt contributing to crust production, which may not necessarily be the same as the volume of melt contributing to outgassing. This study has found stagnant lid extrusive melting rates up to tenfold higher. Therefore melting rates alone would not explain our model's low outgassing fluxes. Hotspots are more geographically sparse, and their global magma supply rate is lower than ridges.}

\editthree{}{\paragraph{Mantle concentrations} The ridge mantle source shows orders-of-magnitude variation in \ch{CO2} concentrations (as derived from melt concentrations). This variation, extending from a minimum of 10~ppm to a maximum of 1980~ppm \citep{Hauri2019}, encompasses our entire range of $\chi_{\ch{CO2}}^{\rm{ini}}$. The hotspot mantle source tends to be even more volatile-rich, up to well over 1000~ppm in \ch{CO2}, and several hundred ppm for \ch{H2O} \citep{Wallace1998, MILLER2019}.}

\editthree{}{\paragraph{Melt concentrations} Our oxidised cases overlap fairly well with estimates of the typical volatile concentrations in modern ridge and hotspot magmas, as well as with the corresponding outgassing fluxes. Simulations that reach higher $\chi_{\ch{CO2}}^{\rm melt}$ and $\chi_{\ch{H2O}}^{\rm melt}$, with respect to modern Earth, accordingly tend to produce higher \ch{CO2} and \ch{H2O} outgassing rates. However, for many of our simulations, the mantle is almost completely depleted in volatiles by 700~Myr. Thus even if the convective velocity is increased by a reduction in viscosity, for example, we would not expect outgassing rates to increase any further.}

\vspace{0.4cm}
\editthree{}{In summary, the most reliable way to attain higher outgassing rates in our stagnant lid model would be to raise the volatile content of the melt. However, a volatile-rich melt is not necessarily attainable by tweaking model parameters because the melt partitioning of \ch{CO2} and \ch{H2O} will be ultimately limited by their stocks in the upper mantle source. To illustrate what a stagnant lid regime's lower interior volatile supplies mean for possible outgassing rates: a mantle with $\chi_{\ch{CO2}}^{\rm ini} = 180$~ppm and no return fluxes could outgas a maximum of $\sim$16 bar \ch{CO2} at 30\% depletion and $\chi_{\rm extr}$ = 40\%. Meanwhile, outgassing 8.5 Tmol~yr$^{-1}$ \ch{CO2} \citep{Catling2017} over 1 Gyr is equivalent to a cumulative $\sim$75 bar.}

\editthree{These figures demonstrate, for instance, how our model tends to produce carbon-poor melts compared to modern magmas \citep{Hauri2019}. Hotspot volcanism in particular has been associated with mantle sources even higher in $f_{\ch{O2}}$, up to the nickle-nickle oxide buffer \citep[e.g.,][]{AMUNDSEN1992, MOUSSALLAM2016}, which is outside the range of graphite stability---making the redox-dependent carbon partitioning model we use here inapplicable. The somewhat-higher integrated melt production compared to hotspots \citep{Mjelde2010} barely compensates for the low \ch{CO2} content, and we ultimately find \ch{CO2} fluxes approaching the lowest estimates of the hotspot outgassing flux \citep{DASGUPTA2010, Hauri2019}. Overall, even for the most fortuitous parameter combinations, outgassing fluxes predicted in this model are at most still several times lower than the modern sum of hotspot, ridge, and plume outgassing (3--10 Tmol yr$^{-1}$ \ch{CO2}) \citep{Gerlach2011}.}{}

\subsubsection{Archean outgassing proxies}

Few observational constraints exist on Archean outgassing rates. Xe isotope anomalies in Archean quartz suggest that mantle degassing was about tenfold greater at 3.3~Ga than at present \citep{Avice2017, martyGeochemicalEvidenceHigh2019}. This proxy would apply to C-O-H outgassing rates at 3.3~Ga if they were derived directly from magma production rates---indeed, these high melting rates suggested by Xe isotopes are matched by our stagnant lid model. However, the carbon and water contents of the melt are at least as important as the volume of the melt in determining C-O-H outgassing rates.

\subsection{Implications for the Archean Earth system}\label{sec:implications}

\subsubsection{From outgassing to atmospheric composition}
\label{sec:atmosphere-discussion}

The atmospheric composition and oxidation state do not mimic the volcanic gas composition or  oxidation state: photochemical reactions and H escape can oxidise reduced gases, and extraterrestrial impactors could provide significant reducing power \citep{Zahnle2020}. In addition, it is possible that an earlier atmosphere degassed from the magma ocean was in some part still present at the end of the Hadean \citep{Hamano2013, Nikolaou_2019, Stueken2020}. We do not model the atmospheric composition here because its complexities deserve more astute attention. \editthree{For completeness, though, this section discusses how several key processes would modulate atmospheric partial pressures post-outgassing.}{However, to build our discussion of this work's potential implications for the Archean Earth system, we will briefly contextualise our results alongside previous work. We focus on processes affecting the partial pressure of \ch{CO2} due to its importance in long-term climate stability.}

\editthree{The equilibrium between outgassing sources and silicate weathering sinks (seafloor and subaerial) controls atmospheric \ch{CO2}. Weathering is thought to regulate the surface temperature via a negative feedback loop \citep{Walker1981}: in the classic \citep{Walker1981} framework, there is a single equilibrium combination of $p$\ch{CO2} and surface temperature for a given outgassing flux.}{}

\editthree{All else being equal, lower outgassing means a cooler climate, a weaker need to weather, and a lower weathering flux\editthree{}{ \citep{Kadoya2014, KT2018}}. \editthree{\citet{KT2018} provide a rare example of a carbon cycle model tested with sub-modern total \ch{CO2} outgassing rates; possible, they acknowledge, with a less-active lid or a low-$f_{\ch{O2}}$ mantle. They}{For example, \citet{KT2018}} find $p$\ch{CO2} as low as $\sim$10$^{-3}$~bar using a minimum outgassing of 1.4 Tmol yr$^{-1}$ \citep{tosi2017habitability}, a flux which is higher than the maximum calculated in this study. At face value, the \citet{KT2018} results might suggest that $p$\ch{CO2} may have been less than $10^{-3}$ bar if the total outgassing were less than 1.4 Tmol yr$^{-1}$. This partial pressure calculation covers a wide range in subaerial land fraction and assumes that weathering is not limited by mineral supply. Following equation (29) in \citet{Foley2018}, the supply limit to silicate weathering would only be reached at \ch{CO2} fluxes of over 100--1000 Tmol yr$^{-1}$ for our stagnant lid scenarios (i.e., it is not reached). For comparison, the earliest likely proxy measurement (3.2~Ga) gives a lower limit of $p$\ch{CO2} $= 2.5 \times 10^{-3}$~bar, based on proxy analyses \citep{Hessler2004}.}{}

\editthree{}{The equilibrium between outgassing sources and silicate weathering sinks (seafloor and subaerial) controls $p$\ch{CO2}. Weathering is thought to regulate the surface temperature via a negative feedback loop: in the classic \citet{Walker1981} framework, there would be a single equilibrium combination of $p$\ch{CO2} and surface temperature for fixed values of the outgassing flux and other relevant parameters (e.g., the weatherable surface area). Lower outgassing---all else being equal---implies a cooler climate, a weaker need to weather, and a lower \ch{CO2} weathering flux\editthree{}{ \citep[e.g.,][]{Kadoya2014, KT2018}}. \editthree{\citet{KT2018} provide a rare example of a carbon cycle model tested with sub-modern total \ch{CO2} outgassing rates; possible, they acknowledge, with a less-active lid or a low-$f_{\ch{O2}}$ mantle. They}{For example, \citet{KT2018}} find $p$\ch{CO2} between 2--500~mbar for a contemporaneous outgassing rate of 3--9~Tmol yr$^{-1}$ (95\% confidence intervals). According to that model, $p$\ch{CO2}~$< 2$~mbar might therefore be anticipated for less than 3~Tmol yr$^{-1}$ of outgassing; such rates are not ruled out by this study. Note that this $p$\ch{CO2} estimate assumes that weathering is not limited by mineral supply, and following equation (29) in \citet{Foley2018}, the supply limit to silicate weathering would only be reached at \ch{CO2} fluxes well over 100~Tmol yr$^{-1}$ in stagnant lid scenarios (i.e., it is not reached). For comparison, proxy analyses by \citet{Hessler2004} give a lower limit of $p$\ch{CO2} $\sim 2.5$~mbar at at 3.2~Ga.}

\subsubsection{Greenhouse warming under the Faint Young Sun}

Despite the fainter luminosity of the young sun, geochemical evidence exists for a temperate climate and stable oceans on Earth as early as 4.4~Ga \citep[e.g.,][]{Wilde2001, Valley2002, Valley2014}. Attempts to explain this paradox often invoke stronger-than-modern atmospheric partial pressures of greenhouse gases \citep[see review in][]{charnay2020}. Volcanic outgassing, being the main source of these greenhouse gases, is a primary control on steady-state $p$\ch{CO2} and surface temperature \citep[e.g.,][]{Walker1981, Sleep2001, Kadoya2014, Honing2019}. \editthree{The additional source of carbonate metamorphism can be discounted in a stagnant lid scenario, assuming that this process primarily occurs in tectonic settings.}{}

Indeed, the \ch{CO2} outgassing rates predicted here are at best \editthree{almost an order of magnitude}{several times} lower than the values often assumed in Archean Earth climate models \citep[e.g.][]{Sleep2001, Wordsworth2013, CHARNAY2017, KANZAKI2018, KT2018}. These higher rates might relate to some combination of a different tectonic mode at the time, a higher ratio of extrusive to intrusive melt, and a more-oxidised mantle. \editthree{Fundamentally, the notion of higher outgassing in the past seems intertwined with the notion of higher mantle heat flow in the past \citep[e.g.,][]{TURCOTTE1980}. As we have argued throughout this study, such a calibration assumes that variations in outgassing are mostly explained by variations in melt production, and that the Archean melt production was necessarily higher.}{} \editthree{Our results would refute both these assumptions under a stagnant lid scenario. Firstly, \ch{CO2} magma concentrations show a stronger correlation with \ch{CO2} outgassing than do melt production rates. Secondly, we nevertheless fail to produce high global melt production rates despite higher mantle temperatures.}{}

\editthree{This work cannot quantify the likelihood of a temperate stagnant lid Archean. Still, a 15\degree C planet warmed purely by \ch{CO2} could be difficult to reconcile with our outgassing model, given the available theory. Namely, 3D general circulation models suggest that maintaining 15\degree C requires 200~mbar of pure \ch{CO2} at 3.8~Ga, if other climate parameters (e.g., day length) are not optimal, and tens of mbar if they are \citep{Wolf2014}.}{3D general circulation models can be used to estimate the minimum partial pressures of \ch{CO2} that would provide the necessary greenhouse warming, for a given solar luminosity. For example, \citet{Wolf2014} find that maintaining 15\degree C requires $\sim$200~mbar of \ch{CO2} at 3.8~Ga, if other climate parameters (e.g., day length) are not optimal, and tens of mbar if they are. Note that this smaller value is similar to the minimum $p$\ch{CO2} from the low-outgassing scenarios studied by \citet{KT2018}.} Dedicated atmospheric modelling should estimate the feasibility of greenhouse warming fed by weaker outgassing in the early Archean. 

At the least, if a planetary mantle $f_{\ch{O2}}$ were below the IW buffer, then it is apparent that virtually no \ch{CO2} could be outgassed in principle, even with melting rates much higher than estimated here. Indeed, silicate Earth may have already reached $f_{\ch{O2}} > {\rm IW} + 1$ by the end of the Hadean \citep{Pahlevan2019}. Depending on the longevity of the primordial magma ocean atmosphere, requiring a \ch{CO2} outgassing rate of at least several Tmol yr$^{-1}$ to sustain an early greenhouse may lend a constraint on the timing of mantle oxidation.

\section{Conclusions}

This work has coupled a 2D numerical model of stagnant lid convection with melting, volatile partitioning into the melt, and chemical speciation of these volatiles. The model returns redox-dependent volcanic outgassing fluxes of \ch{CO2}, \ch{CO}, \ch{H2O}, and \ch{H2}. We have applied this model to Hadean-Archean Earth. We find global \ch{CO2} and \ch{H2O} outgassing fluxes \editthree{of less than $\sim$1 Tmol yr$^{-1}$ and $\sim$}{on the order of 1 Tmol yr$^{-1}$ and }10 Tmol yr$^{-1}$ respectively, depending on the mantle oxidation state. These fluxes are kept low by the assumption of a stagnant lid regime, wherein outgassing may never be much stronger than predicted here if the upper mantle cannot be efficiently replenished in volatiles. Our model may not apply to a scenario where most outgassing occurred on the seafloor, or where the upper mantle were too oxidised for graphite to be the stable form of carbon.

Coupled convection-outgassing models could inform further studies of the early Earth's atmospheric evolution. In particular, unknown \ch{CO2} outgassing rates throughout the late Hadean and early Archean account for a significant portion of the uncertainty on the contemporaneous greenhouse warming capacity. Previous estimates of Earth's climate state during the early period of lower solar luminosity tend to employ outgassing rates higher than the present day, an assumption which might not be substantiated given the unknown tectonic state. If we believe that the presence of surface liquid water in the early Archean demands fairly high \ch{CO2} partial pressures \citep{charnay2020}, then coupled outgassing models might test whether an early initiation of plate tectonics---combined with rapid mantle oxidation---may be key to building a temperate young planet.

\vspace{2cm}

\section*{Acknowledgements}

This manuscript has been markedly improved by the feedback of two anonymous reviewers. Their effort and their insight were critical in every sense, as were the editor's. The authors would like to thank the HPC Service of ZEDAT, Freie Universit\"{a}t Berlin, for computing time. This work was funded by the Deutsche Forschungsgemeinschaft (DFG, German Research Foundation) -- Project-ID 263649064 -- TRR 170. This is TRR 170 Publication No. 142. CMG has received additional funding from the University of Cambridge Harding Distinguished Postgraduate Scholars Programme and the Natural Sciences and Engineering Research Council of Canada (NSERC). Cette recherche a \'{e}t\'{e} financ\'{e}e par le Conseil de recherches en sciences naturelles et en g\'{e}nie du Canada (CRSNG). CMG thanks R. J. Graham and J. Krissansen-Totton for useful discussion, as well as her PhD supervisors O. Shorttle and J. F. Rudge for their patience with this project.

\bibliographystyle{model1-num-names}
\bibliography{references.bib}

\end{document}